\pdfoutput=1
\documentclass[11pt,twoside,a4paper,pdftex,cmspaper,final,collab]{cms-tdr}

\begin{document}\cmsNoteHeader{CFT-09-004}
%%%%%%%%%%%%%%%%%%%%%%%%%%%%%%%%%%%%%%%%%%%%%%%%%%%%%%%%%%%%%%%%%%%%
%
%  Common definitions
%
%  N.B. use of \providecommand rather than \newcommand means
%       that a definition is ignored if already specified
%
%                                              L. Taylor 18 Feb 2005
%%%%%%%%%%%%%%%%%%%%%%%%%%%%%%%%%%%%%%%%%%%%%%%%%%%%%%%%%%%%%%%%%%%%

%%%%%%%%%%%%%%%%%%%%%%%%%%%%%%%%%%%%%%%%%%%%%%%%%%%%%%%%%%%%%%%%%%%%
%
% Hyphenations (only need to add here if you get a nasty word break)
%
\hyphenation{env-iron-men-tal}%    just an example
\hyphenation{had-ron-i-za-tion}
\hyphenation{cal-or-i-me-ter}
\hyphenation{de-vices}
%
% Hyphenations-end
%
% CVS info. These are modified by cvs at checkout time.
% The last version of these macros found before the maketitle will be the one on the front page,
% so only the main file is tracked.
% Edit by hand with care!
\RCS$Revision: 1.80 $
\RCS$Date: 2009/12/23 23:03:31 $
\RCS$Name:  $
%%%%%%%%%%%%% ptdr definitions %%%%%%%%%%%%%%%%%%%%%
%%%%%%%%%%%%%%%%%%%%%%%%%%%%%%%%%%%%%%%%%%%%%%%%%%%%%%%%%%%%%%%%%%%%
%
%  Common definitions
%
%  N.B. use of \providecommand rather than \newcommand means
%       that a definition is ignored if already specified
%
%                                              L. Taylor 18 Feb 2005
%%%%%%%%%%%%%%%%%%%%%%%%%%%%%%%%%%%%%%%%%%%%%%%%%%%%%%%%%%%%%%%%%%%%

% Some shorthand
% turn off italics
\newcommand {\etal}{\mbox{et al.}\xspace} %et al. - no preceding comma
\newcommand {\ie}{\mbox{i.e.}\xspace}     %i.e.
\newcommand {\eg}{\mbox{e.g.}\xspace}     %e.g.
\newcommand {\etc}{\mbox{etc.}\xspace}     %etc.
\newcommand {\vs}{\mbox{\sl vs.}\xspace}      %vs.
\newcommand {\mdash}{\ensuremath{\mathrm{-}}} % for use within formulas

% some terms whose definition we may change
\newcommand {\Lone}{Level-1\xspace} % Level-1 or L1 ?
\newcommand {\Ltwo}{Level-2\xspace}
\newcommand {\Lthree}{Level-3\xspace}

% Some software programs (alphabetized)
\providecommand{\ACERMC} {\textsc{AcerMC}\xspace}
\providecommand{\ALPGEN} {{\textsc{alpgen}}\xspace}
\providecommand{\CHARYBDIS} {{\textsc{charybdis}}\xspace}
\providecommand{\CMKIN} {\textsc{cmkin}\xspace}
\providecommand{\CMSIM} {{\textsc{cmsim}}\xspace}
\providecommand{\CMSSW} {{\textsc{cmssw}}\xspace}
\providecommand{\COBRA} {{\textsc{cobra}}\xspace}
\providecommand{\COCOA} {{\textsc{cocoa}}\xspace}
\providecommand{\COMPHEP} {\textsc{CompHEP}\xspace}
\providecommand{\EVTGEN} {{\textsc{evtgen}}\xspace}
\providecommand{\FAMOS} {{\textsc{famos}}\xspace}
\providecommand{\GARCON} {\textsc{garcon}\xspace}
\providecommand{\GARFIELD} {{\textsc{garfield}}\xspace}
\providecommand{\GEANE} {{\textsc{geane}}\xspace}
\providecommand{\GEANTfour} {{\textsc{geant4}}\xspace}
\providecommand{\GEANTthree} {{\textsc{geant3}}\xspace}
\providecommand{\GEANT} {{\textsc{geant}}\xspace}
\providecommand{\HDECAY} {\textsc{hdecay}\xspace}
\providecommand{\HERWIG} {{\textsc{herwig}}\xspace}
\providecommand{\HIGLU} {{\textsc{higlu}}\xspace}
\providecommand{\HIJING} {{\textsc{hijing}}\xspace}
\providecommand{\IGUANA} {\textsc{iguana}\xspace}
\providecommand{\ISAJET} {{\textsc{isajet}}\xspace}
\providecommand{\ISAPYTHIA} {{\textsc{isapythia}}\xspace}
\providecommand{\ISASUGRA} {{\textsc{isasugra}}\xspace}
\providecommand{\ISASUSY} {{\textsc{isasusy}}\xspace}
\providecommand{\ISAWIG} {{\textsc{isawig}}\xspace}
\providecommand{\MADGRAPH} {\textsc{MadGraph}\xspace}
\providecommand{\MCATNLO} {\textsc{mc@nlo}\xspace}
\providecommand{\MCFM} {\textsc{mcfm}\xspace}
\providecommand{\MILLEPEDE} {{\textsc{millepede}}\xspace}
\providecommand{\ORCA} {{\textsc{orca}}\xspace}
\providecommand{\OSCAR} {{\textsc{oscar}}\xspace}
\providecommand{\PHOTOS} {\textsc{photos}\xspace}
\providecommand{\PROSPINO} {\textsc{prospino}\xspace}
\providecommand{\PYTHIA} {{\textsc{pythia}}\xspace}
\providecommand{\SHERPA} {{\textsc{sherpa}}\xspace}
\providecommand{\TAUOLA} {\textsc{tauola}\xspace}
\providecommand{\TOPREX} {\textsc{TopReX}\xspace}
\providecommand{\XDAQ} {{\textsc{xdaq}}\xspace}

%  Experiments
\newcommand {\DZERO}{D\O\xspace}     %etc.

% Measurements and units...

\newcommand{\de}{\ensuremath{^\circ}}
\newcommand{\ten}[1]{\ensuremath{\times \text{10}^\text{#1}}}
\newcommand{\unit}[1]{\ensuremath{\text{\,#1}}\xspace}
\newcommand{\mum}{\ensuremath{\,\mu\text{m}}\xspace}
\newcommand{\micron}{\ensuremath{\,\mu\text{m}}\xspace}
\newcommand{\cm}{\ensuremath{\,\text{cm}}\xspace}
\newcommand{\mm}{\ensuremath{\,\text{mm}}\xspace}
\newcommand{\mus}{\ensuremath{\,\mu\text{s}}\xspace}
\newcommand{\keV}{\ensuremath{\,\text{ke\hspace{-.08em}V}}\xspace}
\newcommand{\MeV}{\ensuremath{\,\text{Me\hspace{-.08em}V}}\xspace}
\newcommand{\GeV}{\ensuremath{\,\text{Ge\hspace{-.08em}V}}\xspace}
\newcommand{\TeV}{\ensuremath{\,\text{Te\hspace{-.08em}V}}\xspace}
\newcommand{\PeV}{\ensuremath{\,\text{Pe\hspace{-.08em}V}}\xspace}
\newcommand{\keVc}{\ensuremath{{\,\text{ke\hspace{-.08em}V\hspace{-0.16em}/\hspace{-0.08em}c}}}\xspace}
\newcommand{\MeVc}{\ensuremath{{\,\text{Me\hspace{-.08em}V\hspace{-0.16em}/\hspace{-0.08em}c}}}\xspace}
\newcommand{\GeVc}{\ensuremath{{\,\text{Ge\hspace{-.08em}V\hspace{-0.16em}/\hspace{-0.08em}c}}}\xspace}
\newcommand{\TeVc}{\ensuremath{{\,\text{Te\hspace{-.08em}V\hspace{-0.16em}/\hspace{-0.08em}c}}}\xspace}
\newcommand{\keVcc}{\ensuremath{{\,\text{ke\hspace{-.08em}V\hspace{-0.16em}/\hspace{-0.08em}c}^\text{2}}}\xspace}
\newcommand{\MeVcc}{\ensuremath{{\,\text{Me\hspace{-.08em}V\hspace{-0.16em}/\hspace{-0.08em}c}^\text{2}}}\xspace}
\newcommand{\GeVcc}{\ensuremath{{\,\text{Ge\hspace{-.08em}V\hspace{-0.16em}/\hspace{-0.08em}c}^\text{2}}}\xspace}
\newcommand{\TeVcc}{\ensuremath{{\,\text{Te\hspace{-.08em}V\hspace{-0.16em}/\hspace{-0.08em}c}^\text{2}}}\xspace}

\newcommand{\pbinv} {\mbox{\ensuremath{\,\text{pb}^\text{$-$1}}}\xspace}
\newcommand{\fbinv} {\mbox{\ensuremath{\,\text{fb}^\text{$-$1}}}\xspace}
\newcommand{\nbinv} {\mbox{\ensuremath{\,\text{nb}^\text{$-$1}}}\xspace}
\newcommand{\percms}{\ensuremath{\,\text{cm}^\text{$-$2}\,\text{s}^\text{$-$1}}\xspace}
\newcommand{\lumi}{\ensuremath{\mathcal{L}}\xspace}
\newcommand{\Lumi}{\ensuremath{\mathcal{L}}\xspace}%both upper and lower
%
% Need a convention here:
\newcommand{\LvLow}  {\ensuremath{\mathcal{L}=\text{10}^\text{32}\,\text{cm}^\text{$-$2}\,\text{s}^\text{$-$1}}\xspace}
\newcommand{\LLow}   {\ensuremath{\mathcal{L}=\text{10}^\text{33}\,\text{cm}^\text{$-$2}\,\text{s}^\text{$-$1}}\xspace}
\newcommand{\lowlumi}{\ensuremath{\mathcal{L}=\text{2}\times \text{10}^\text{33}\,\text{cm}^\text{$-$2}\,\text{s}^\text{$-$1}}\xspace}
\newcommand{\LMed}   {\ensuremath{\mathcal{L}=\text{2}\times \text{10}^\text{33}\,\text{cm}^\text{$-$2}\,\text{s}^\text{$-$1}}\xspace}
\newcommand{\LHigh}  {\ensuremath{\mathcal{L}=\text{10}^\text{34}\,\text{cm}^\text{$-$2}\,\text{s}^\text{$-$1}}\xspace}
\newcommand{\hilumi} {\ensuremath{\mathcal{L}=\text{10}^\text{34}\,\text{cm}^\text{$-$2}\,\text{s}^\text{$-$1}}\xspace}

% Some usual physics terms

\newcommand{\zp}{\ensuremath{\mathrm{Z}^\prime}\xspace}

% SM (still to be classified)

\newcommand{\kt}{\ensuremath{k_{\mathrm{T}}}\xspace}
\newcommand{\BC}{\ensuremath{{B_{\mathrm{c}}}}\xspace}
\newcommand{\bbarc}{\ensuremath{{\overline{b}c}}\xspace}
\newcommand{\bbbar}{\ensuremath{{b\overline{b}}}\xspace}
\newcommand{\ccbar}{\ensuremath{{c\overline{c}}}\xspace}
\newcommand{\JPsi}{\ensuremath{{J}/\psi}\xspace}
\newcommand{\bspsiphi}{\ensuremath{B_s \to \JPsi\, \phi}\xspace}
\newcommand{\AFB}{\ensuremath{A_\mathrm{FB}}\xspace}
\newcommand{\EE}{\ensuremath{e^+e^-}\xspace}
\newcommand{\MM}{\ensuremath{\mu^+\mu^-}\xspace}
\newcommand{\TT}{\ensuremath{\tau^+\tau^-}\xspace}
\newcommand{\wangle}{\ensuremath{\sin^{2}\theta_{\mathrm{eff}}^\mathrm{lept}(M^2_\mathrm{Z})}\xspace}
\newcommand{\ttbar}{\ensuremath{{t\overline{t}}}\xspace}
\newcommand{\stat}{\ensuremath{\,\text{(stat.)}}\xspace}
\newcommand{\syst}{\ensuremath{\,\text{(syst.)}}\xspace}
% these moved to similar defs
%\newcommand{\Etmiss}{\ensuremath{E_{\mathrm{T}\!{\rm miss}}}}
%\newcommand{\VEtmiss}{\ensuremath{{\vec E}_{\mathrm{T}\!{\rm miss}}}}

%%%  E-gamma definitions
\newcommand{\HGG}{\ensuremath{\mathrm{H}\to\gamma\gamma}}
\newcommand{\gev}{\GeV}
\newcommand{\GAMJET}{\ensuremath{\gamma + \mathrm{jet}}}
\newcommand{\PPTOJETS}{\ensuremath{\mathrm{pp}\to\mathrm{jets}}}
\newcommand{\PPTOGG}{\ensuremath{\mathrm{pp}\to\gamma\gamma}}
\newcommand{\PPTOGAMJET}{\ensuremath{\mathrm{pp}\to\gamma +
\mathrm{jet}
}}
\newcommand{\MH}{\ensuremath{\mathrm{M_{\mathrm{H}}}}}
\newcommand{\RNINE}{\ensuremath{\mathrm{R}_\mathrm{9}}}
\newcommand{\DR}{\ensuremath{\Delta\mathrm{R}}}

% Physics symbols ...

\newcommand{\PT}{\ensuremath{p_{\mathrm{T}}}\xspace}
\newcommand{\pt}{\ensuremath{p_{\mathrm{T}}}\xspace}
\newcommand{\ET}{\ensuremath{E_{\mathrm{T}}}\xspace}
\newcommand{\HT}{\ensuremath{H_{\mathrm{T}}}\xspace}
\newcommand{\et}{\ensuremath{E_{\mathrm{T}}}\xspace}
\newcommand{\Em}{\ensuremath{E\!\!\!/}\xspace}
\newcommand{\Pm}{\ensuremath{p\!\!\!/}\xspace}
\newcommand{\PTm}{\ensuremath{{p\!\!\!/}_{\mathrm{T}}}\xspace}
\newcommand{\ETm}{\ensuremath{E_{\mathrm{T}}^{\mathrm{miss}}}\xspace}
\newcommand{\MET}{\ensuremath{E_{\mathrm{T}}^{\mathrm{miss}}}\xspace}
\newcommand{\ETmiss}{\ensuremath{E_{\mathrm{T}}^{\mathrm{miss}}}\xspace}
\newcommand{\VEtmiss}{\ensuremath{{\vec E}_{\mathrm{T}}^{\mathrm{miss}}}\xspace}

%%%%%%
% From Albert
%

\newcommand{\ga}{\ensuremath{\gtrsim}}
\newcommand{\la}{\ensuremath{\lesssim}}
\newcommand{\swsq}{\ensuremath{\sin^2\theta_W}\xspace}
\newcommand{\cwsq}{\ensuremath{\cos^2\theta_W}\xspace}
\newcommand{\tanb}{\ensuremath{\tan\beta}\xspace}
\newcommand{\tanbsq}{\ensuremath{\tan^{2}\beta}\xspace}
\newcommand{\sidb}{\ensuremath{\sin 2\beta}\xspace}
\newcommand{\alpS}{\ensuremath{\alpha_S}\xspace}
\newcommand{\alpt}{\ensuremath{\tilde{\alpha}}\xspace}

\newcommand{\QL}{\ensuremath{Q_L}\xspace}
\newcommand{\sQ}{\ensuremath{\tilde{Q}}\xspace}
\newcommand{\sQL}{\ensuremath{\tilde{Q}_L}\xspace}
\newcommand{\ULC}{\ensuremath{U_L^C}\xspace}
\newcommand{\sUC}{\ensuremath{\tilde{U}^C}\xspace}
\newcommand{\sULC}{\ensuremath{\tilde{U}_L^C}\xspace}
\newcommand{\DLC}{\ensuremath{D_L^C}\xspace}
\newcommand{\sDC}{\ensuremath{\tilde{D}^C}\xspace}
\newcommand{\sDLC}{\ensuremath{\tilde{D}_L^C}\xspace}
\newcommand{\LL}{\ensuremath{L_L}\xspace}
\newcommand{\sL}{\ensuremath{\tilde{L}}\xspace}
\newcommand{\sLL}{\ensuremath{\tilde{L}_L}\xspace}
\newcommand{\ELC}{\ensuremath{E_L^C}\xspace}
\newcommand{\sEC}{\ensuremath{\tilde{E}^C}\xspace}
\newcommand{\sELC}{\ensuremath{\tilde{E}_L^C}\xspace}
\newcommand{\sEL}{\ensuremath{\tilde{E}_L}\xspace}
\newcommand{\sER}{\ensuremath{\tilde{E}_R}\xspace}
\newcommand{\sFer}{\ensuremath{\tilde{f}}\xspace}
\newcommand{\sQua}{\ensuremath{\tilde{q}}\xspace}
\newcommand{\sUp}{\ensuremath{\tilde{u}}\xspace}
\newcommand{\suL}{\ensuremath{\tilde{u}_L}\xspace}
\newcommand{\suR}{\ensuremath{\tilde{u}_R}\xspace}
\newcommand{\sDw}{\ensuremath{\tilde{d}}\xspace}
\newcommand{\sdL}{\ensuremath{\tilde{d}_L}\xspace}
\newcommand{\sdR}{\ensuremath{\tilde{d}_R}\xspace}
\newcommand{\sTop}{\ensuremath{\tilde{t}}\xspace}
\newcommand{\stL}{\ensuremath{\tilde{t}_L}\xspace}
\newcommand{\stR}{\ensuremath{\tilde{t}_R}\xspace}
\newcommand{\stone}{\ensuremath{\tilde{t}_1}\xspace}
\newcommand{\sttwo}{\ensuremath{\tilde{t}_2}\xspace}
\newcommand{\sBot}{\ensuremath{\tilde{b}}\xspace}
\newcommand{\sbL}{\ensuremath{\tilde{b}_L}\xspace}
\newcommand{\sbR}{\ensuremath{\tilde{b}_R}\xspace}
\newcommand{\sbone}{\ensuremath{\tilde{b}_1}\xspace}
\newcommand{\sbtwo}{\ensuremath{\tilde{b}_2}\xspace}
\newcommand{\sLep}{\ensuremath{\tilde{l}}\xspace}
\newcommand{\sLepC}{\ensuremath{\tilde{l}^C}\xspace}
\newcommand{\sEl}{\ensuremath{\tilde{e}}\xspace}
\newcommand{\sElC}{\ensuremath{\tilde{e}^C}\xspace}
\newcommand{\seL}{\ensuremath{\tilde{e}_L}\xspace}
\newcommand{\seR}{\ensuremath{\tilde{e}_R}\xspace}
\newcommand{\snL}{\ensuremath{\tilde{\nu}_L}\xspace}
\newcommand{\sMu}{\ensuremath{\tilde{\mu}}\xspace}
\newcommand{\sNu}{\ensuremath{\tilde{\nu}}\xspace}
\newcommand{\sTau}{\ensuremath{\tilde{\tau}}\xspace}
\newcommand{\Glu}{\ensuremath{g}\xspace}
\newcommand{\sGlu}{\ensuremath{\tilde{g}}\xspace}
\newcommand{\Wpm}{\ensuremath{W^{\pm}}\xspace}
\newcommand{\sWpm}{\ensuremath{\tilde{W}^{\pm}}\xspace}
\newcommand{\Wz}{\ensuremath{W^{0}}\xspace}
\newcommand{\sWz}{\ensuremath{\tilde{W}^{0}}\xspace}
\newcommand{\sWino}{\ensuremath{\tilde{W}}\xspace}
\newcommand{\Bz}{\ensuremath{B^{0}}\xspace}
\newcommand{\sBz}{\ensuremath{\tilde{B}^{0}}\xspace}
\newcommand{\sBino}{\ensuremath{\tilde{B}}\xspace}
\newcommand{\Zz}{\ensuremath{Z^{0}}\xspace}
\newcommand{\sZino}{\ensuremath{\tilde{Z}^{0}}\xspace}
\newcommand{\sGam}{\ensuremath{\tilde{\gamma}}\xspace}
\newcommand{\chiz}{\ensuremath{\tilde{\chi}^{0}}\xspace}
\newcommand{\chip}{\ensuremath{\tilde{\chi}^{+}}\xspace}
\newcommand{\chim}{\ensuremath{\tilde{\chi}^{-}}\xspace}
\newcommand{\chipm}{\ensuremath{\tilde{\chi}^{\pm}}\xspace}
\newcommand{\Hone}{\ensuremath{H_{d}}\xspace}
\newcommand{\sHone}{\ensuremath{\tilde{H}_{d}}\xspace}
\newcommand{\Htwo}{\ensuremath{H_{u}}\xspace}
\newcommand{\sHtwo}{\ensuremath{\tilde{H}_{u}}\xspace}
\newcommand{\sHig}{\ensuremath{\tilde{H}}\xspace}
\newcommand{\sHa}{\ensuremath{\tilde{H}_{a}}\xspace}
\newcommand{\sHb}{\ensuremath{\tilde{H}_{b}}\xspace}
\newcommand{\sHpm}{\ensuremath{\tilde{H}^{\pm}}\xspace}
\newcommand{\hz}{\ensuremath{h^{0}}\xspace}
\newcommand{\Hz}{\ensuremath{H^{0}}\xspace}
\newcommand{\Az}{\ensuremath{A^{0}}\xspace}
\newcommand{\Hpm}{\ensuremath{H^{\pm}}\xspace}
\newcommand{\sGra}{\ensuremath{\tilde{G}}\xspace}
\newcommand{\mtil}{\ensuremath{\tilde{m}}\xspace}
\newcommand{\rpv}{\ensuremath{\rlap{\kern.2em/}R}\xspace}
\newcommand{\LLE}{\ensuremath{LL\bar{E}}\xspace}
\newcommand{\LQD}{\ensuremath{LQ\bar{D}}\xspace}
\newcommand{\UDD}{\ensuremath{\overline{UDD}}\xspace}
\newcommand{\Lam}{\ensuremath{\lambda}\xspace}
\newcommand{\Lamp}{\ensuremath{\lambda'}\xspace}
\newcommand{\Lampp}{\ensuremath{\lambda''}\xspace}
\newcommand{\spinbd}[2]{\ensuremath{\bar{#1}_{\dot{#2}}}\xspace}

\newcommand{\MD}{\ensuremath{{M_\mathrm{D}}}\xspace}% ED mass
\newcommand{\Mpl}{\ensuremath{{M_\mathrm{Pl}}}\xspace}% Planck mass
\newcommand{\Rinv} {\ensuremath{{R}^{-1}}\xspace}

%%%%%%%%%%%%%%%%%%%%%%%%%%%%%%%%%%%%%%%%%%%%%%%%%%%%%%%%%%%%%%%%%%%%
%
% Hyphenations (only need to add here if you get a nasty word break)
%
\hyphenation{en-viron-men-tal}%    just an example

%%%%%%%%%%%%%%%  Title page %%%%%%%%%%%%%%%%%%%%%%%%
\cmsNoteHeader{09-004}
%\title{Crystal ECAL Performance and Operation}% Force line breaks with \\
\title{Performance and Operation of the CMS Electromagnetic Calorimeter}% Force line breaks with \\

%Author is always "The CMS Collaboration" for PAS, so author, etc will be ignored
\address[cern]{CERN}
\author[cern]{The CMS Collaboration}

% please supply the date in yyyy/mm/dd format. Today has been
% redefined to do so, but it should be fixed as of the final release date.
\date{\today}

% note that you cannot use \verb in the abstract text
\abstract{
  The operation and general performance of the CMS electromagnetic calorimeter using cosmic-ray muons are described. These muons were recorded after the closure of the CMS detector in late 2008. The calorimeter is made of lead tungstate crystals and the overall status of the 75\,848 channels corresponding to the barrel and endcap detectors is reported. The stability of crucial operational parameters, such as high voltage, temperature and electronic noise, is summarised and the performance of  the light monitoring system is presented.    
}

% these need to be filled in by hand and should (MUST) match the info
% in the TeX equivalents less the TeX markup
\hypersetup{%
pdfauthor={The CMS Collaboration},%
pdftitle={Performance and Operation of the CMS Electromagnetic Calorimeter},%
pdfsubject={CMS},%
pdfkeywords={CMS, ECAL, electromagnetic calorimeter, performance, CRAFT}}

\maketitle %maketitle comes after all the front information has been supplied

%%pdftitle={Crystal ECAL Performance and Operation},%

%%%%%%%%%%%%%%%%%%%%%%%%%%%%%%%%  Begin text %%%%%%%%%%%%%%%%%%%%%%%%%%%%%
\section{Introduction}

The primary goal of the Compact Muon Solenoid (CMS) experiment~\cite{:2008zzk} is to explore physics at the TeV energy scale, exploiting the proton-proton collisions delivered by the Large Hadron Collider (LHC)~\cite{lhc}. The main component of the CMS detector to identify and measure photons and electrons is the electromagnetic calorimeter (ECAL)~\cite{:2008zzk,CERN_LHCC_97-033}. The CMS ECAL is %%a fine grained crystal calorimeter~\cite{annrev} 
designed with stringent requirements on energy resolution, in order to be sensitive to the decay of a Higgs boson into two photons.

\begin{figure}[hbtp]
  \begin{center}
    \includegraphics[width=0.8\textwidth]{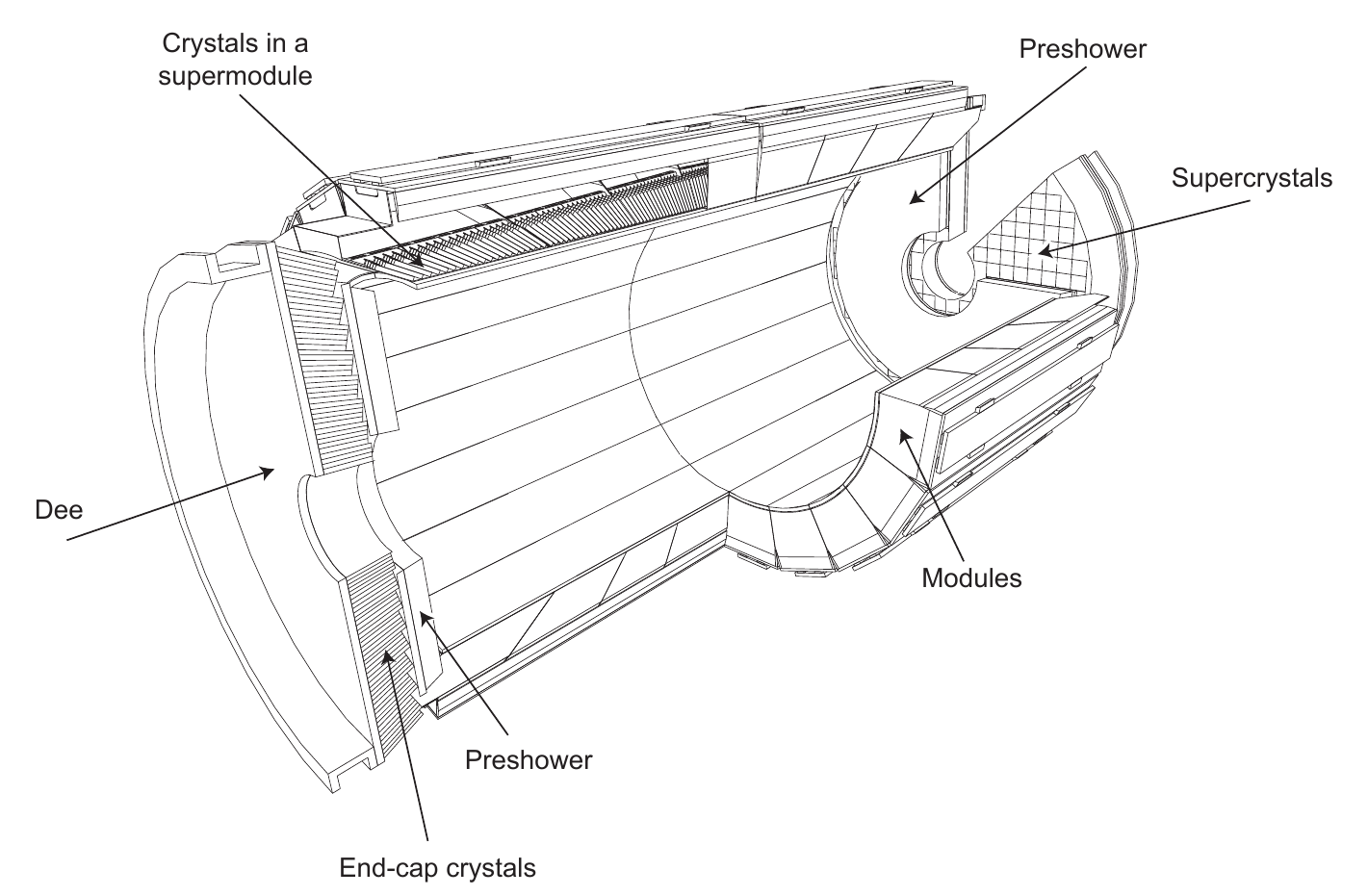}
    \caption{Layout of the CMS electromagnetic calorimeter, showing the barrel supermodules, the two endcaps and the preshower detectors.}
    \label{fig:ecal}
  \end{center}
\end{figure}

Crystal calorimeters have the potential to provide fast response, radiation tolerance and excellent energy resolution~\cite{annrev}. The CMS ECAL is composed of 75\,848 lead tungstate (PbWO$_4$) crystals. The detector consists of a barrel region,
%% constructed from 36 individual supermodules, 
extending to a pseudorapidity $|\eta|$ of 1.48, and two endcaps, which extend coverage to $|\eta|=3.0$. Scintillation light from the crystals is detected by avalanche photodiodes (APDs) in the barrel region and by vacuum phototriodes (VPTs) in the endcaps. The layout of the CMS ECAL, showing the crystal barrel and endcap detectors, as well as the silicon preshower detectors, is shown in Fig.~\ref{fig:ecal}.

In order to achieve the desired energy resolution of the ECAL it is 
necessary to maintain the stability of the per-channel energy calibration over time. 
This places stringent requirements on the stability of the temperature of the ECAL and of the high voltage applied to the APDs. This is due to the temperature dependence of the crystal light yield, as well as the sensitivity of the APD gains to variations in both temperature and high voltage (the VPT response is much less sensitive to temperature and high voltage variations). In addition, changes in crystal transparency under irradiation must be tracked and corrected for.

During October-November 2008 the CMS Collaboration conducted a month-long data taking exercise known as Cosmic Run At Four Tesla (CRAFT), with  the goal of commissioning the experiment for an extended operating period~\cite{CRAFTGeneral}. With all installed detector systems participating, CMS recorded 270 million cosmic-ray muon events with the solenoid at its nominal axial magnetic field strength of 3.8~T. These tests were the first opportunity to exercise over an extended period of time the electromagnetic calorimeter as installed within CMS. The performance results from the ECAL during these tests are reported in this paper.

The paper is structured as follows. Section~\ref{sec:two} provides a brief description of the ECAL and summarises the installation of the barrel and endcaps in CMS. Sections~\ref{sec:three}-\ref{sec:six} deal mostly with the analysis of data recorded by the ECAL during CRAFT. Section~\ref{sec:three} describes the algorithms used to reconstruct the energy deposited in the detector by cosmic-ray muons. Section~\ref{sec:four} shows the achieved stability of temperature, high voltage and electronic noise. These measurements are compared to the stability levels needed in order to achieve the desired energy resolution performance of the ECAL. Progress in validating the light monitoring system is also described in this section. Section~\ref{sec:calib} presents the results from the use of cosmic-ray and beam-induced muons (the latter from LHC operation in September 2008) to verify the pre-existing calibration constants, which were obtained from laboratory and test beam measurements. In Section~\ref{sec:six}, the results from a series of dedicated calibration runs that were taken in the endcap detectors are described. These runs were used to make measurements of the VPT response in the 3.8~T CMS magnetic field, in order to update the existing endcap calibration constants that were obtained at zero magnetic field, and to measure the effect of pulsing rate on VPT stability.

\section{The ECAL in CMS}\label{sec:two}

Each of the 36 supermodules in the ECAL barrel (EB) consists of 1700 tapered PbWO$_4$ crystals with a frontal area of approximately $2.2\times2.2$~cm$^2$  and a length of 23~cm (corresponding to 25.8 radiation lengths). The crystal axes are inclined at an angle of 3$^{\circ}$ relative to the direction of  the nominal interaction point, in both the azimuthal ($\phi$) and $\eta$ projections. Scintillation light from the crystals  is detected by two Hamamatsu S8148 $5\times5$~mm$^2$ APDs (approximately 4.5 photoelectrons per MeV at 18~$^{\circ}$C), which were specially developed for CMS and operate at a gain of 50. These are connected in parallel to the on-detector readout electronics, which are organised in units of $5\times5$ crystals, each unit corresponding to a trigger tower. Each trigger tower consists of five Very Front End (VFE) cards, each accepting data from 5 APD pairs. The APD signals are pre-amplified and shaped by Multiple Gain Pre-Amplifier (MGPA) ASICs located on the VFE boards, which consist of three parallel amplification stages (gains 1, 6 and 12)~\cite{CERN_LHCC_2003-055}. The output is digitised by a 12-bit ADC running at 40 MHz, which samples the pulse ten times for each channel and selects the gain with the highest non-saturated signal. The data from five VFE cards are transferred to a single front-end card, which generates the trigger primitive data~\cite{Paganini:2009zz}, and transmits it to the dedicated off-detector trigger electronics.

The two ECAL endcaps (EE) are constructed from four half-disk `dees', each consisting of 3662 tapered crystals, with a frontal area of $2.68\times 2.68$~cm$^2$ and a length of 22~cm (corresponding to 24.7 radiation lengths), arranged in a quasi-projective geometry. The crystals are focussed at a point 1.3~m farther than the nominal interaction point along the beam line, with off-pointing angles between $2^{\circ}$ and $8^{\circ}$. The crystals in each dee are organised into 138 standard $5\times 5$ supercrystal units, and 18 special shaped supercrystals that are located at the inner and outer radii. Scintillation light is detected by  VPTs (type PMT188) produced by NRIE with an active area of 280~mm$^2$ and operating at gains of 8--10, which are glued to the rear face of the crystals.
The VPTs installed in CMS have a  25\% (RMS) spread in anode sensitivity and were sorted into six batches across the detector. The highest sensitivity VPTs are installed along the outer circumference of the endcaps and the lowest sensitivity tubes are installed along the inner circumference, ensuring a roughly constant transverse energy equivalent of the noise as a function of $\eta$.   Further details of the design and construction of the ECAL, the associated on-detector and off-detector readout electronics, and the performance of individual system components can be found elsewhere~\cite{:2008zzk}.

Installation of the ECAL barrel into CMS was performed during 2007. The last module was installed in July of that year and the integration of essential detector services (low voltage, high voltage and cooling) and preliminary commissioning of the supermodules was completed in December 2007. Prior to this, all supermodules were fully tested in the laboratory after construction and were exposed to cosmic-ray muons for a period of ten days, to obtain relative channel-to-channel inter-calibration constants. Nine of the 36 supermodules were also exposed to test beam electrons to provide absolute energy calibrations (described further in Section~\ref{sec:calib}). During 2006, two supermodules were installed and tested in the CMS solenoid at 4~T along with other sub-detectors~\cite{4179147}.
The endcap dees were constructed and commissioned at CERN during early 2008. The dees were installed in CMS during July 2008 and the entire barrel and endcap calorimeter was commissioned prior to the closure of CMS in late August, in preparation for first LHC beam. The silicon preshower detectors, which are located in front of the ECAL endcaps, were not included in CMS for the 2008 run. They were installed during early 2009 and have been fully commissioned prior to LHC operation in late 2009.

\section{ECAL operation during CRAFT}\label{sec:three}

Of the 270 million cosmic-ray events recorded by CMS at 3.8~T during CRAFT, a total of 246 million were used in ECAL reconstruction and analysis. Of these, 158 million events were taken with the nominal APD gain of 50 (G50), in order to study trigger performance, noise and the signatures of minimum ionising particles (MIP) in the configuration that will be used for collision data. In order to study cosmic-ray muon signatures in ECAL with greater efficiency, the remaining 88 million events were taken with APD gain 200 (G200). 
For these two gains, the average ADC to MeV conversion factors in EB are 1~ADC count $\approx38$ $(9.3)$~MeV for G50 (G200). As discussed in Section~\ref{sec:noise}, the single channel noise is unchanged in G50 and G200, leading to an increased signal to noise ratio for cosmic-ray muons in G200. 

The ECAL trigger was operated in the barrel region during CRAFT, using data taken with APD G50. The trigger algorithm used in CRAFT, which is described in detail in Ref.~\cite{Paganini:2009zz}, involves the generation of trigger primitive data for each trigger tower. These provide a measurement of the total transverse energy ($E_T$) of the trigger tower, as well as a single (``fine grain'') bit that indicates a compact lateral extent of the energy deposit. In CRAFT, a threshold on the trigger primitive $E_T$ of 750~MeV was applied at the trigger tower level. These trigger primitives are sent to the Regional Calorimeter Trigger (RCT)~\cite{CERN_LHCC_2000-038}. Electromagnetic candidates were formed by requiring that the summed $E_T$ in two neighbouring towers exceeds a threshold of 1~GeV\footnote{A much higher threshold on the $E_T$ of electromagnetic candidates will be applied for LHC beam running. For a luminosity of $2\times 10^{33}$~cm$^{-2}$s$^{-1}$, a threshold of 26~GeV is envisaged for single electron/photon candidates~\cite{CERN_LHCC_2006-021}.}, that the fine grain bit is set, and that the associated energy deposition in the hadronic calorimeter is low relative to the energy deposited in the ECAL ($<5\%$). The typical trigger rate during CRAFT was 30--40~Hz. Further details can be found in Ref.~\cite{ecaltrig}. 

A data reduction algorithm, termed selective readout~\cite{Almeida:2005aa}, is applied to reduce the ECAL raw data size to the level of 100~kB/event, which is the bandwidth allocated to the calorimeter readout by the CMS DAQ system. During CRAFT, the trigger towers in EB for a particular event were classified as low or high interest, based on their measured $E_T$.  Towers in EB with $E_{T}$ greater than 687.5~MeV (APD G50) were classified as of high interest. For low interest towers ($E_{T}<687.5$~MeV), only channels with amplitude above a minimum threshold, termed the zero suppression threshold, were read out. All channels in a $3\times 3$ trigger tower matrix centred on a high interest tower were read out. The zero suppression threshold was 2.25 ADC counts (approximately 90~MeV in G50), corresponding to approximately twice the measured noise level in the highest MGPA gain. In EE, a zero suppression threshold of 3.0 ADC counts (1.5 times the noise level in the highest MGPA gain) was applied to all channels.

A fit was performed to the 10 digitised 25~ns time samples surrounding a signal, in order to estimate the signal amplitude and timing for each channel that is read out. The delay of the readout pipeline is such that the signal pulse is expected to start from the fourth sample and the baseline pedestal value can be estimated from the first three digitised samples, termed pre-samples~\cite{Adzic:2006nn}.  

 Two different amplitude reconstruction algorithms are used in this paper. For the analysis of cosmic-ray muons (described in Section~\ref{sec:calib}), which are asynchronous with respect to the 40~MHz sampling frequency of the ADC, a parameterised pulse shape function was used, with fixed shape parameters optimised separately for barrel and endcap crystals. This algorithm was also used in the analyses described in Section~\ref{sec:las} and in Section~\ref{sec:six}.

For LHC beam running, where the readout samples will be synchronised to the 40~MHz LHC frequency, the standard amplitude reconstruction method is a digital filtering technique~\cite{Adzic:2006nn}. This method estimates the pulse amplitude by a linear weighting of the individual samples, and requires that the position of the pulse maximum has a small jitter (within 1~ns~\cite{Adzic:2006nn}). For the analyses presented in this paper the signal amplitude was reconstructed using five consecutive digitised samples around the expected position of the peak and dynamically subtracting the pedestal from each event using the three pre-samples before the peak.
 This ``3+5 weight'' algorithm was used, as described in Section~\ref{sec:noise}, to estimate the electronic noise  that would be obtained from the amplitude reconstruction method intended for LHC running. Further details of the ECAL time reconstruction methods, and their performance during CRAFT, can be found in Ref.~\cite{ecaltime}.

The reconstructed hits for each event are grouped into clusters of $5\times 5$ crystals. At CRAFT, the clusters were seeded from a single crystal with a reconstructed amplitude greater than 15 ADC counts above pedestal (corresponding to 130~MeV for APD G200 or 12.5 standard deviations above the noise in the highest MGPA gain), or from two adjacent crystals with amplitudes greater than 5 ADC counts (approximately 40~MeV in G200) above pedestal. Contiguous clusters are grouped together to form superclusters in order to collect the energy deposited by muons which traverse the ECAL at large angles with respect to the crystal axes.
With APD G200, the probability for a muon traversing the length of a crystal in the ECAL barrel to produce a reconstructed cluster was greater than 99\%~\cite{ecaldedx}. The analysis described in Section~\ref{sec:dedx} uses only muons that cross the tracker volume. For these muons, which should pass through the ECAL twice, there was an average of 1.7 reconstructed superclusters per event. The reduction from the expected number of two is due to some muons either passing through temporarily non-operating regions of the barrel (supermodules with low voltage turned off) or passing through the forward regions of the detector.

The fraction of channels that were operational during CRAFT was 98.33\% in EB (60177/61200) and 99.66\% in EE (14598/14648).  For the barrel, 28/2448 trigger towers (1.14\%) were turned off due to a damaged low voltage supply cable, which was repaired after CRAFT. The readout of 11 trigger towers (0.45\%) was suppressed due to data integrity problems. A total of 48 channels (0.08\%) were classified as inoperable, based on pedestal, charge injection and laser calibration measurements, as well as beam-induced muon data from the September 2008 LHC beam tests. An additional 35 single channels were classified as problematic, but could be operated in at least one of the three MGPA gains. These  83 (48 inoperable plus 35 problematic) channels were masked in the ECAL cosmic-ray muon reconstruction. Most of them have been known since detector commissioning, with 16 new channels discovered since the installation in CMS. For the endcaps, data from two supercrystals, corresponding to 50/14648 channels (0.34\%) were suppressed due to a broken data optical fibre inside one of the dees (25 channels) and a faulty low voltage connection powering five VFEs (25 channels). No isolated dead channel was observed in the endcap data.

\section{System stability}\label{sec:four}

The electromagnetic (EM) energy resolution of ECAL can be parameterised as a function of the incident electron/photon energy, E (in GeV), as

\begin{equation}
\frac{\sigma_{E}}{E}=\frac{a}{\sqrt{E}}\oplus\frac{b}{E}\oplus c\quad ,\label{eq:energyres}
\end{equation}

where $a$ represents the stochastic term, which depends on event to event fluctuations in lateral shower containment, photo-statistics and photodetector gain; $b$ represents the noise term, which depends on the level of electronic noise and event pile-up (additional particles causing signals that overlap in time); and $c$ represents the constant term, which depends on non-uniformity of the longitudinal light collection, leakage of energy from the rear face of the crystal and the accuracy of the detector inter-calibration constants. The target value for the constant term, which dominates the resolution at high energies, is 0.5\% for both the barrel and the endcaps~\cite{CERN_LHCC_97-033}. 

Previous measurements taken with test beam electrons with energies between 20 and 250~GeV have shown that the EM energy resolution and noise performance of the ECAL barrel meets the design goals for the detector. For the barrel, the mean values of the stochastic and constant terms, computed using the energy summed over 3x3 crystal arrays, are $2.8\%/\sqrt{E({\rm GeV})}$ and 0.3\% respectively~\cite{Adzic:2007mi}. The mean single channel noise, computed for 1175 crystals, is 41.5 MeV energy equivalent. The following sections describe the achieved stability of electronic noise (Section~\ref{sec:noise}), high voltage (Section~\ref{sec:hv}), temperature (Section~\ref{sec:temp}), and the ECAL light monitoring system (Section~\ref{sec:las}) for data taken during 2008. The purpose of these measurements is to show that the operating conditions during CRAFT meet the ECAL goals on detector stability, and that the observed high voltage and temperature fluctuations provide a negligible contribution to the constant term of the EM energy resolution. 
 
The stability of these quantities was monitored using data collected in dedicated runs, as well as data taken continuously throughout CRAFT in the ECAL calibration sequence.  This sequence periodically injected pedestal, MGPA test pulse (charge injection to the front-end electronics) and laser events into the data stream during the simulated LHC abort gap (3~$\mu$s gap at the end of each 89~$\mu$s beam cycle).

\subsection{Noise stability}\label{sec:noise}

\begin{figure}[hbtp]
  \begin{center}
    \includegraphics[width=0.5\textwidth]{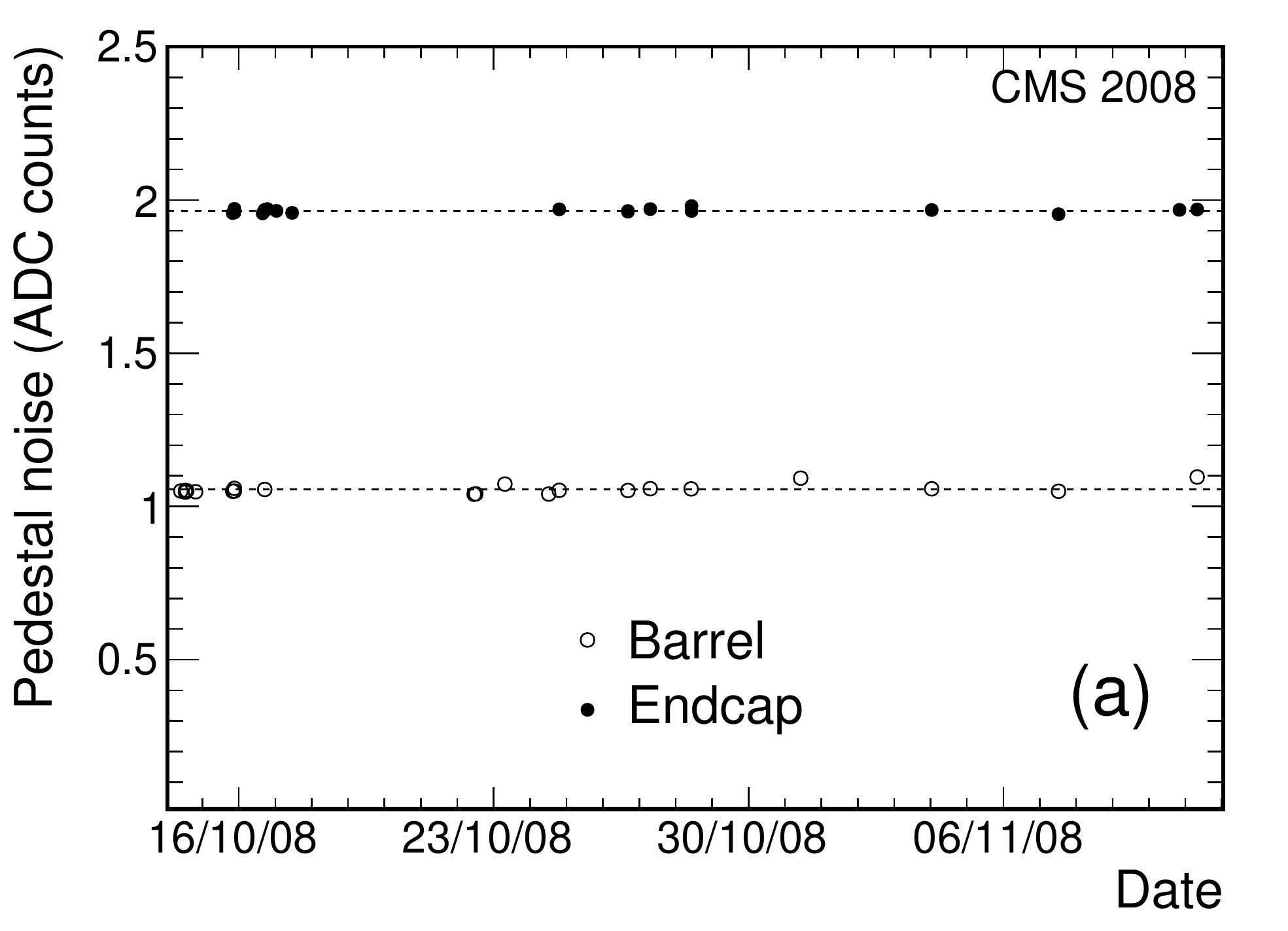}\includegraphics[width=0.5\textwidth]{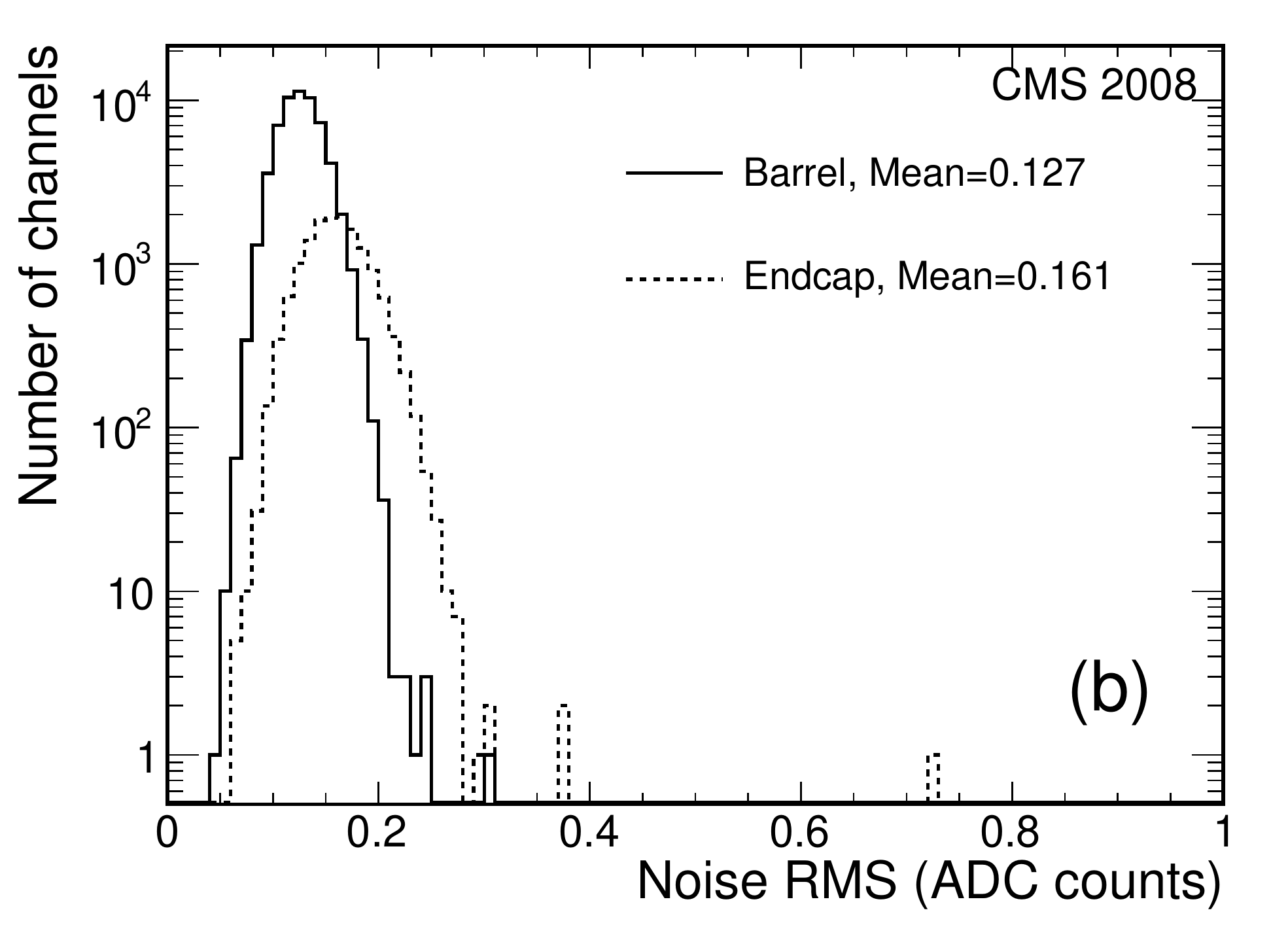}
    \caption{Pedestal noise stability during CRAFT. (a) Average electronic noise in the barrel (open circles) and endcaps (filled circles) versus time for pedestal data taken during CRAFT. Data in the highest MGPA gain are used, and the noise level is expressed in ADC counts.  The dashed lines represent the average noise levels in the barrel and endcaps, over the time period shown. (b) Distribution of the RMS of the pedestal noise for each channel, measured in the pedestal runs shown in (a), plotted separately for barrel and endcap channels.}
    \label{fig:noise}
  \end{center}
\end{figure}

The electronic noise of the ECAL was monitored during CRAFT from dedicated pedestal runs, which measure the noise in all three gains of the MGPA in the absence of signal pulses.
Figures~\ref{fig:noise}(a) and (b) show the stability in EB and EE of the per-channel noise level (expressed in ADC counts) for the highest MGPA gain, which is the most sensitive to electronic noise.

 For barrel data, the ``3+5 weight'' amplitude reconstruction method was used to estimate the level of electronic noise. As stated above, this is the baseline method for clock-synchronous LHC running and, since the three pre-samples are used to subtract the pedestal for each event, it is known to be effective in reducing the level of low frequency (or pickup) noise~\cite{Adzic:2006nn}. An additional noise contribution, observed in 9 of the 36 barrel supermodules during CRAFT and believed to be low frequency pickup noise ($<4$~MHz) associated with the operation of other CMS sub-detectors in the underground cavern, was observed in the individual time samples.   This excess noise was strongly suppressed by the digital filtering technique. Variations in the mean noise level per supermodule of 25\% were reduced to less than 2\%, consistent with statistical uncertainties. The noise level was defined as the RMS deviation of the reconstructed signal amplitude measured from each pedestal event. The noise level is uniform across all barrel supermodules following the application of this method.

For the endcap detectors no significant source of pickup noise was observed. Accordingly, it was not necessary to apply the digital filtering amplitude reconstruction method to obtain good noise performance and stability. For endcap data, the noise level was defined as the RMS deviation of the three pre-samples, summed over all pedestal events.

The data points shown in Fig.~\ref{fig:noise}(a) come from several different runs, taken with the CMS magnetic field at 0~T or 3.8~T and in the barrel with APD gains G50 or G200. The fact that all measurements are perfectly aligned shows that the noise level in the ECAL does not depend on the CMS magnetic field nor on the APD gain. For the highest MGPA gain, the average noise level per channel was 1.06 ADC counts in the barrel, and 1.96 ADC counts in the endcaps.   The observed noise levels in EB and EE during CRAFT are consistent with the values measured during module construction (see, for example, Table 1 of Ref.~\cite{Adzic:2006nn} for EB noise measurements obtained with test beam data using the ``3+5 weight'' method), and meet the MGPA design specifications~\cite{CERN_LHCC_2003-055}. For the barrel, the average value of the noise in energy equivalent units corresponds to roughly 40~MeV/channel (APD G50).

A small number ($<0.1\%$) of single channels showed high noise levels during CRAFT, either high pedestal RMS (greater than 2.0 ADC counts in the barrel and greater than 3.0 ADC counts in the endcaps) or high occupancy in cosmic-ray muon runs. These channels were excluded in the subsequent reprocessing of the CRAFT data.  The per-channel noise stability during CRAFT is shown in Fig.~\ref{fig:noise}(b). It shows the RMS of the variations of the noise levels measured in the highest MGPA gain for each individual channel, and computed over the pedestal runs used in Fig.~\ref{fig:noise}(a). The average per-channel variation was 0.127 ADC counts in the barrel and 0.161 ADC counts in the endcaps.  The performance of the MGPA was also shown to be insensitive to the CMS magnetic field at better than the per mille level, using dedicated charge injection runs.

\subsection{High voltage stability}\label{sec:hv}

High voltage (HV) is supplied to the barrel APDs via a custom HV power supply developed in collaboration with CAEN. A total of 18 CAEN SY1527 crates are used. These are located in the CMS service cavern at a distance of 120~m from the ECAL, and sense wires are used to correct for voltage drops in the HV supply lines between the crates and the detector. Each crate contains eight A1520E boards, which carry up to 9 channels. Each channel can provide a bias voltage of 0--500~V to 50 APD pairs with a maximum current of 15~mA. A total of 1224 HV channels are used in the ECAL barrel. The APDs are sorted according to operating voltage, and paired such that the mean gain is 50.  The nominal operating voltage is between 340 and 430~V. Since the APD gain, $G$, is very sensitive to the bias voltage, $1/G \; (\partial G/\partial V)\approx 3\%/$V, the operating voltage must be kept stable to better than 60~mV to provide a negligible contribution to the constant term of the EM energy resolution. The HV crates are fully integrated into the CMS Detector Control System (DCS) framework, which allows the applied voltages and currents for each channel to be remotely controlled and monitored.
High voltage is supplied to the endcap VPTs by two CAEN SY1527 crates, one for each endcap. The cathodes are at ground potential, the dynodes are held at 600~V and the anodes at 800~V. One pair of CAEN channels (one for anodes, one for dynodes) serves approximately one quadrant (four pairs at each endcap). In addition, there is an interlock on the CAEN boards, to switch off the high voltage to the VPTs if the CMS magnetic field is not at a constant value. At the operating bias used in CMS, the VPT gain is close to saturation~\cite{:2008zzk}. As a result, the voltages for the endcaps do not have to be controlled very precisely (the VPT gain dependence on high voltage is less than 0.1$\%/$V~\cite{Bell:2004eu}).

\begin{figure}[hbtp]
  \begin{center}
    \includegraphics[width=0.5\textwidth]{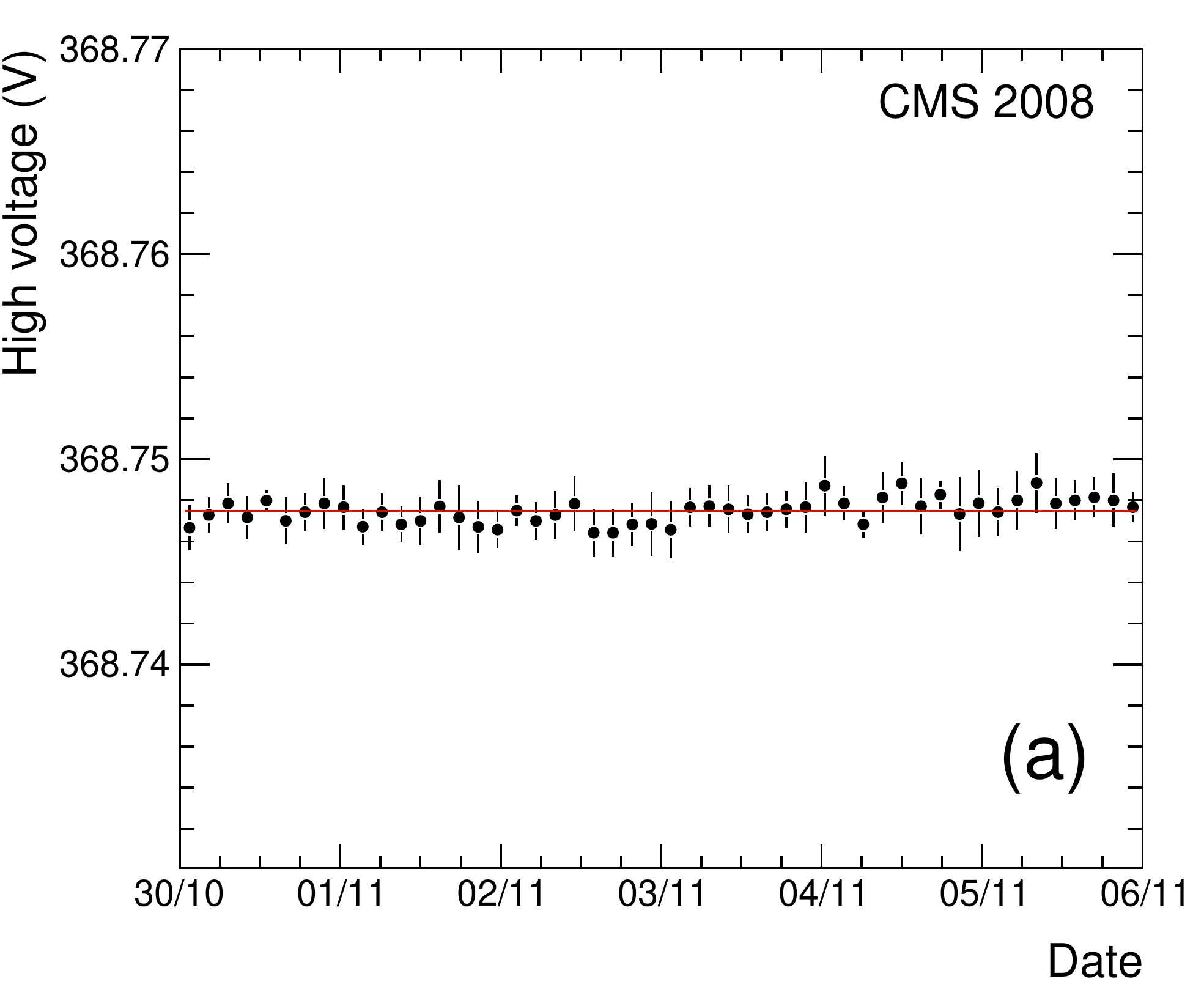}\includegraphics[width=0.5\textwidth]{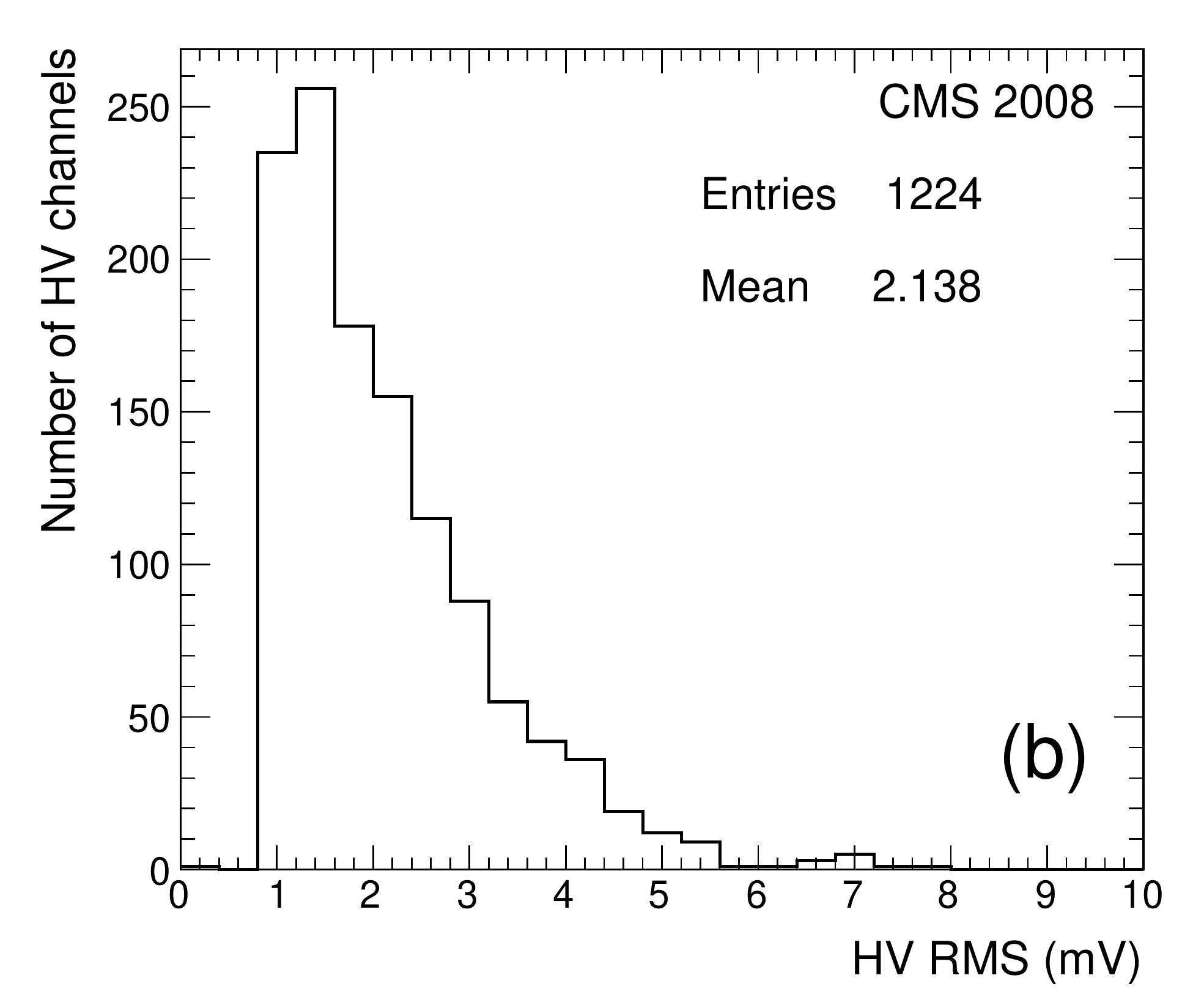}
    \caption{Barrel HV stability during CRAFT. (a)
    Monitored HV on a barrel sense wire during one week of CRAFT
    data taking when the APD gain was set to G50. Each data point is averaged over a three hour time period. 
    (b) Distribution of the RMS of the readings for each HV channel during this period.}
    \label{fig:EB_HV_stability} 
    \end{center}
\end{figure}

During the CRAFT data taking period, high voltage was supplied to the barrel APDs with two
different settings, corresponding to G50 and G200.  For the purpose of measuring high voltage stability, a one-week period during CRAFT has been selected,
when all channels were continuously operated at APD G50. The typical current drawn by each HV channel during this period 
was 2--3~$\mu$A.  Figure~\ref{fig:EB_HV_stability}(a) 
shows the monitored voltage on
one HV sense wire, as recorded by the CAEN crate and logged in the CMS detector conditions
database~\cite{craftdataflow}.  All the points are compatible with a constant value within
the measurement errors.  The line represents the average over this period.
The stability of the sense wire readings for the barrel HV channels during this period can be estimated by the
distribution of the RMS of the readings of each individual channel (Fig.~\ref{fig:EB_HV_stability}(b)).
The average fluctuation of the high voltage is 2.1~mV (RMS).  More than 97\% of the total number of channels have fluctuations below 5~mV and all were within 10~mV during the
time period considered here.

APD dark current measurements were recorded for each channel by Detector Control Unit (DCU) ASICs located on the front-end electronics. The additional voltage drop over the 136~k$\Omega$ protection resistor between the sense point and the APD cathode could have a sizeable effect on the applied voltage for leakage currents of a fraction of a $\mu$A.
 The minimum dark current measurable by the DCU system, once the DCU readout pedestal has been subtracted, is 0.32~$\mu$A. The ADC pedestals have been computed averaging several runs taken with no high voltage applied to the APDs. The measurements recorded during the CRAFT data taking reported dark currents below the measurable threshold for almost all channels in the barrel, as expected for non-irradiated APDs, only 11 channels ($<0.02\%$) showing measurable currents.

\subsection{Temperature stability}\label{sec:temp}

The temperature of the ECAL barrel is required to be stable within 0.05~$^{\circ}$C. This ensures that temperature fluctuations provide a negligible contribution to the constant term of the EM energy resolution. Fluctuations in temperature directly affect the light yield of the crystals (the temperature dependence of the light yield is approximately $-2\%$ per $^{\circ}$C) and the gain of the APDs in the ECAL barrel, $1/G \; (\partial G/\partial T)\approx-2.3\%/^{\circ}$C~\cite{Renker:2002st}. In the endcaps, the temperature dependence of the VPT response is assumed to be negligible relative to the temperature sensitivity of the crystal light yield~\cite{et1,et2}. Accordingly, a less stringent temperature stability requirement of 0.1~$^{\circ}$C is assumed for the endcap dees.

The nominal operating temperature of ECAL is 18~$^{\circ}$C. A cooling system utilising water flow~\cite{:2008zzk,CERN_LHCC_97-033} is used to regulate the temperature of the barrel and endcap crystals, which are thermally decoupled from the silicon tracker and preshower detectors. In addition, the return water is distributed to a series of aluminium cooling bars, which are coupled to the very front end electronics and remove the heat generated by these components. 

Temperature readings are provided by two independent groups of sensors. Precision Temperature Monitor (PTM) devices (10 per supermodule, 24 per endcap dee) measure the temperatures on each side of the crystal volume and the incoming and outgoing cooling water. These are a set of precision temperature sensors (NTC 470~$\Omega$ thermistors manufactured by EPCOS) read out via CAN-bus, which have a relative accuracy of $\approx 0.01~^{\circ}$C. In addition, thermistors are fixed to the back of each $5\times 2$ crystal matrix (170 per supermodule) in the barrel and to each supercrystal in the endcaps. These thermistors were read out by the DCU ASICs located on each VFE board. The thermistors were calibrated in the laboratory prior to installation, and the response of the DCU ASICs was then calibrated by the PTM devices.

\begin{figure}[hbtp]
  \begin{center}
    \includegraphics[width=0.8\textwidth]{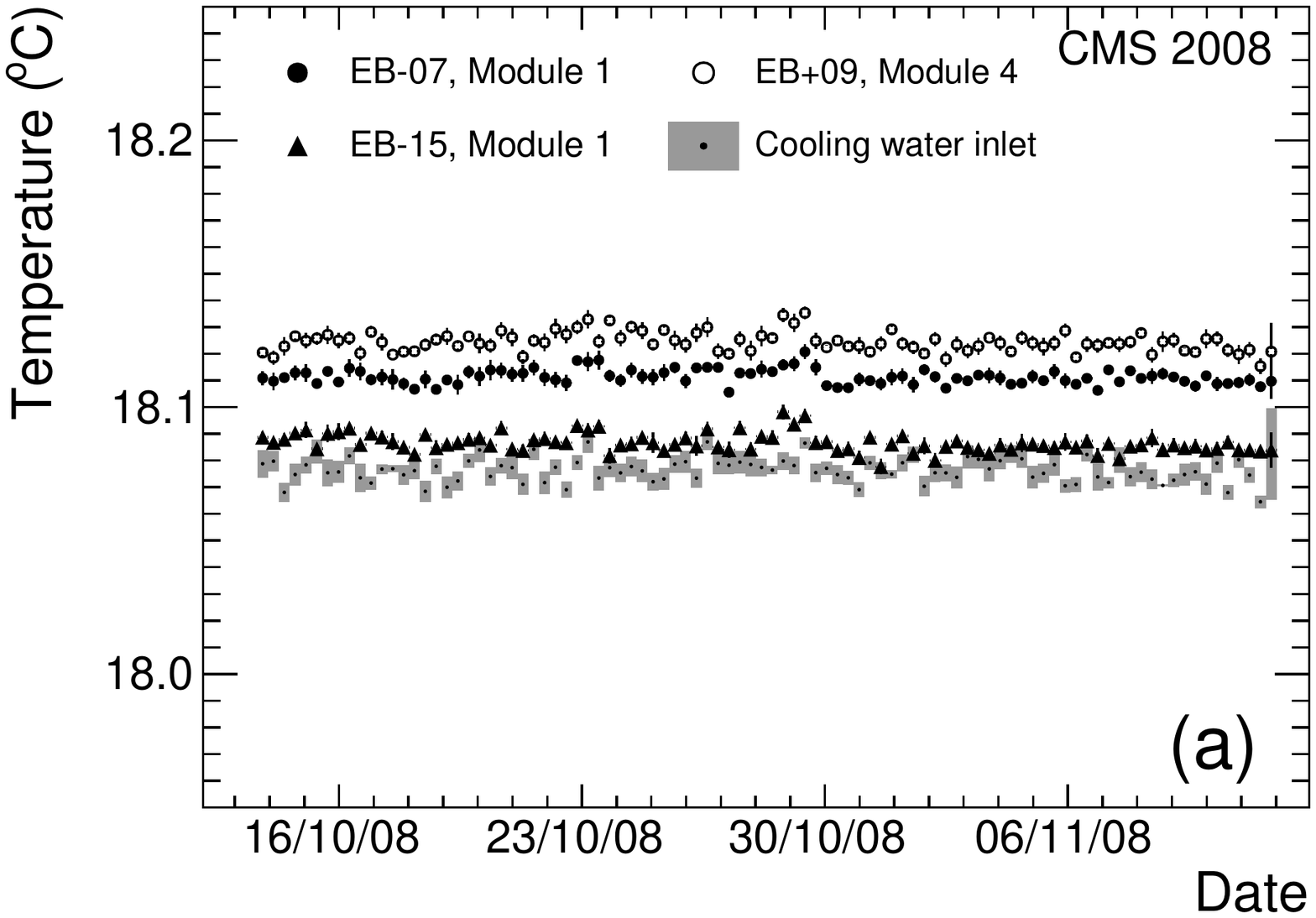}
    \includegraphics[width=0.8\textwidth]{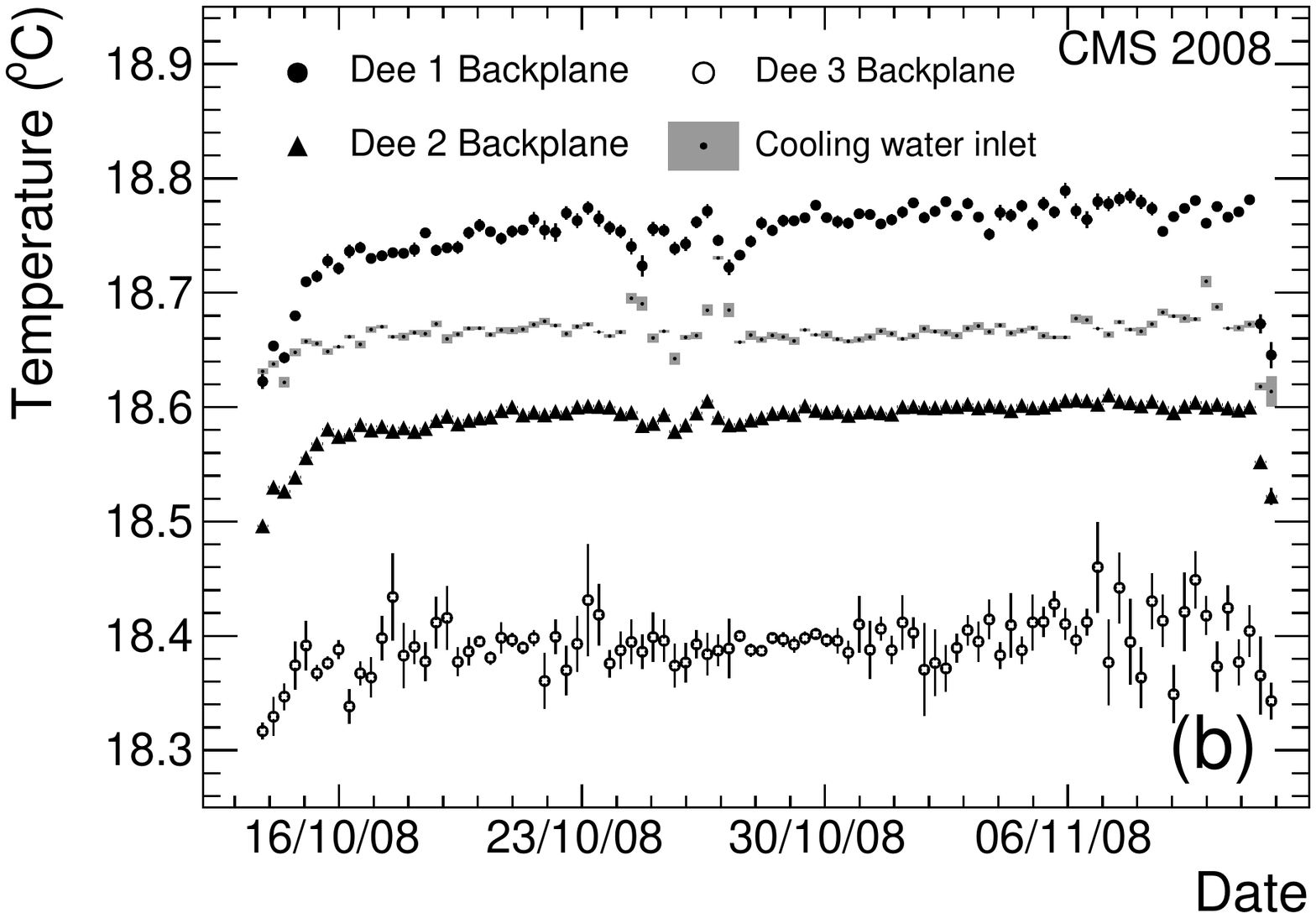}
    \caption{Stability of ECAL temperature during CRAFT.  (a) Mean temperatures recorded over eight-hour time bins for PTM sensors located in four different ECAL barrel supermodules. Three of the four sensors monitored the temperature close to the rear face of the crystals, and one sensor recorded the input water temperature in one of the cooling circuits. (b) Mean temperatures recorded for four representative PTM sensors in the endcap. Three sensors monitored the temperature on the dee backplanes, and one monitored the input cooling water temperature. The error bars represent the error on the mean of approximately 45 measurements per data point.}
    \label{fig:eb_temp}
  \end{center}
\end{figure}

Figure~\ref{fig:eb_temp}(a) shows the EB temperature history during CRAFT for three representative PTM sensors, monitoring the temperature close to the rear face of the crystals of three different supermodules. Measurements from an additional sensor, monitoring the input cooling water temperature of another supermodule, are also shown. Each data point is the average of approximately 45 readings taken over an eight-hour period, and the error bar  represents the uncertainty on this mean value. During CRAFT, these temperature readings were stable to better than 0.01~$^{\circ}$C, which is well within the desired stability target. Temperature sensors in both the innermost ($|\eta|<0.44$) and outermost ($1.13<|\eta|<1.48$) regions of the supermodules are shown (labelled Module 1 and Module 4, respectively). The outer regions of the supermodules are observed to be hotter than the inner regions by 0.09~$^{\circ}$C, on average. This is probably due to the fact that the former are close to the supermodule patch panels, where all services, cooling manifolds and cables converge. The mean temperature measured in the ECAL barrel during CRAFT was $18.10\pm 0.02~^{\circ}$C by the PTM sensors and $18.12\pm 0.04$~$^{\circ}$C by the APD capsule thermistors. 

Figure~\ref{fig:eb_temp}(b) shows the EE temperature history during CRAFT. Three representative PTM sensors are shown, reading temperatures on the dee backplates, close to the rear face of the crystals. An additional sensor monitoring the input cooling water is also shown for reference. The readings are shown to be stable within $\pm 0.02$~$^{\circ}$C after October 15th, following an initial period of temperature stabilization. This is well within the ECAL requirement for the temperature stability of the endcap detectors. The observed small changes in the backplane sensor readings are correlated with fluctuations in the input water temperature. The readings in dee 3 clearly fluctuate much more than those of the other sensors, during much of the CRAFT running period. Comparing the general patterns, it is seen that the fluctuations in the dee 3 data are due to noise in the sensor and not to temperature instabilities. The mean temperature measured by the PTM sensors in the two endcaps during CRAFT was $18.58\pm0.03~^{\circ}$C for dee 1 and dee 2, and $18.55\pm0.06~^{\circ}$C for dee 3 and dee 4. The larger RMS in the second endcap is caused by the higher noise level observed in the dee 3 PTM sensors. The PTM temperature profiles were examined for data taken after CRAFT in order to investigate the slow rise in temperatures observed in Fig.~\ref{fig:eb_temp}(b). No evidence of long-term temperature drifts was seen.

The RMS deviation of temperature histories was also calculated for the 6009 barrel thermistors and  548 endcap thermistors that were read out during CRAFT. The average stability was 0.009~$^{\circ}$C in the barrel, with all measurements within the ECAL specification of 0.05~$^{\circ}$C. The average stability in the endcaps was measured to be 0.017~$^{\circ}$C, using data from 15$^{\mathrm{th}}$ October onwards, once the temperature had stabilised following the initial turn-on period clearly visible in Fig.~\ref{fig:eb_temp}(b).  Measurements comparing the variation of neighbouring thermistors in the barrel and endcaps indicated a higher level of readout noise in the latter. However, even if all of the observed fluctuations in the endcap thermistor readings are attributed to temperature instabilities, practically all of the measurements lie within the specification of 0.1~$^{\circ}$C.

\subsection{Crystal transparency monitoring}\label{sec:las}

The ECAL laser monitoring (LM) system~\cite{CMS_NOTE_2007-028} is critical for maintaining the stability of the constant term of the EM energy resolution at high luminosities. Its main purpose is to accurately measure and to correct for changes in the lead tungstate crystal transparency, which will decrease during irradiation at the LHC due to formation of colour centres. The crystals will slowly recover transparency through annealing when beams are off. The LM system is also able to detect and correct for other effects such as photodetector gain changes due to temperature or high voltage variations.

To reach the ECAL design performance, the LM system is required to monitor transparency changes for each crystal at the $0.2\%$ level, with one measurement every 20 to 30 minutes. The LM system consists of two different lasers: a blue laser with a wavelength (440~nm) close to the emission peak of scintillation light from PbWO$_4$, and an infra-red laser (wavelength~796 nm) for which crystal transparency is stable under irradiation. The blue laser is used to monitor crystal transparency to scintillation light whereas the infra-red laser is used to disentangle effects due to irradiation from other possible effects such as gain variations.

Light is fanned out from the laser sources to the 75\,848 crystals by means of a two-level distribution system. A fibre optic switch directs laser pulses from the laser source located in the CMS service cavern via optical fibres to a single calorimeter element located on the detector. There are 72 half-supermodule calorimeter elements in the barrel and 16 quarter-dee elements in the endcaps. The secondary fanout consists of a reflective light splitter, and 9 (19) output optical fibres per barrel supermodule (endcap dee). The tertiary fanout consists of a 12~mm inner-diameter thermoplastic light diffusing sphere with a fanout of typically 200 fibres that carry the laser light to individual crystals.  Laboratory measurements indicate a typical spread in light yield of 2.4\% (RMS) over 240 fibres. For the endcaps, this tertiary light distribution system is shared with a LED pulser system, which was installed in 2008 when the endcap dees were installed in their final position, in the experimental cavern. The LED system contains 76 light sources in two colours: blue (455~nm) and orange (617~nm). Its main purpose is to provide a constant background pulsing rate of $>100$~Hz to mitigate the effect of VPT anode sensitivity to the rate, as described in Section~6.2. Additional fanout fibres are connected to a set of 528 radiation-hard PN diodes, which provide monitoring of the laser and LED light output, and allow pulse-to-pulse variations in the reconstructed amplitudes to be corrected for.

Changes in the crystal transparency due to radiation damage do not, however, affect the amplitude from the APD signal for an electromagnetic shower (S) in exactly the same way as they affect the signal for injected laser pulses (R), due to the different mean path lengths of the light in the crystals. It has been shown that it is possible to relate the signals in the two cases simply by: $\frac{S}{S_0}=\big ( \frac{R}{R_0} \big )^{\alpha}$. This expression, with $\alpha \approx 1.6$, was shown to describe well the behaviour of crystals evaluated using test beam data~\cite{:2008zzk,Adzic:2006za}.
   
During CRAFT, a total of approximately 500 sequences of laser monitoring data were taken within the ECAL calibration sequence. The laser typically ran at 100~Hz, resulting in the injection of laser light into $\mathcal{O}(1\%)$ of the available LHC beam abort gaps. 

\begin{figure}[hbtp]
  \begin{center}
    \includegraphics[width=0.5\textwidth]{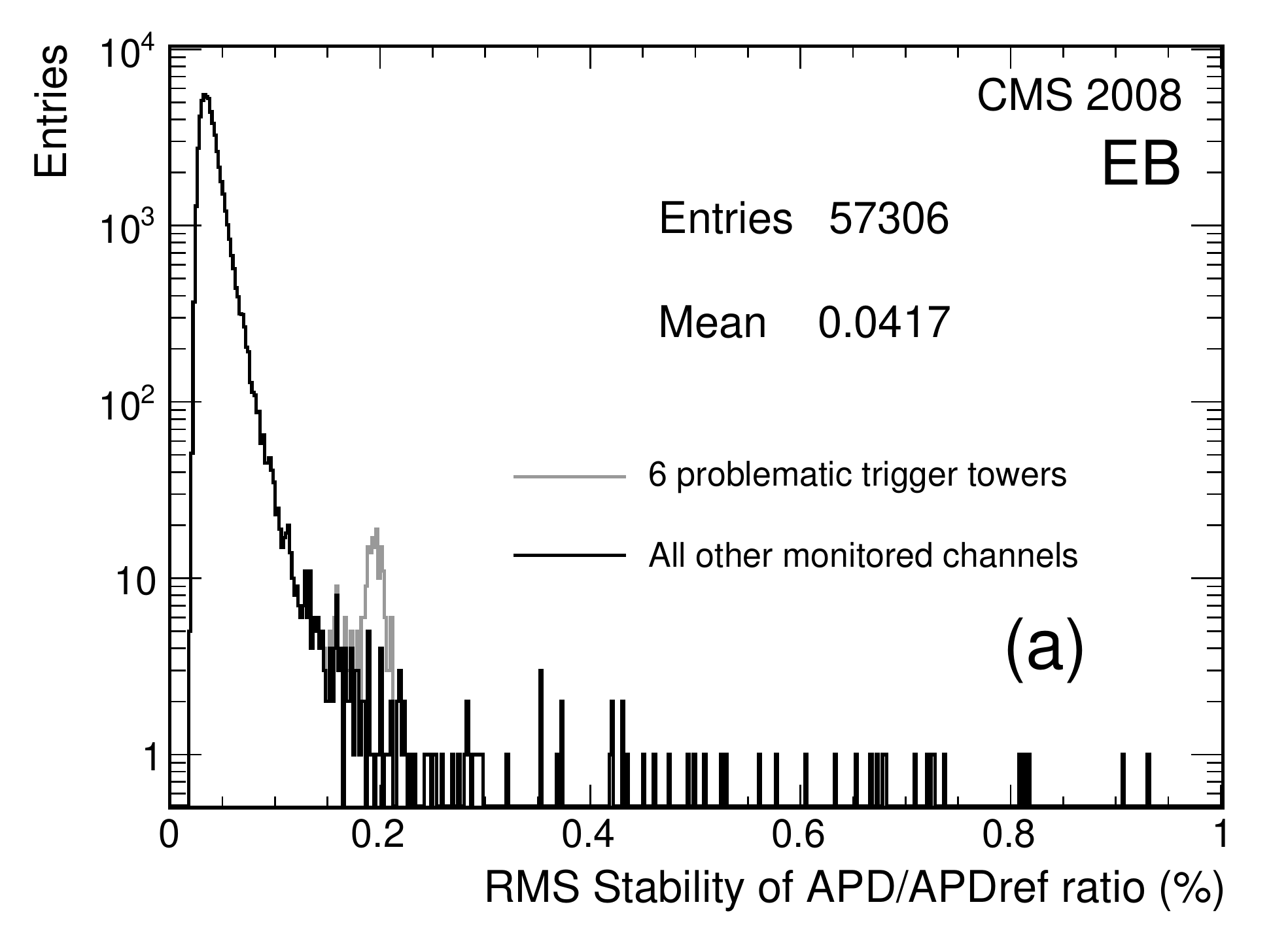}\includegraphics[width=0.5\textwidth]{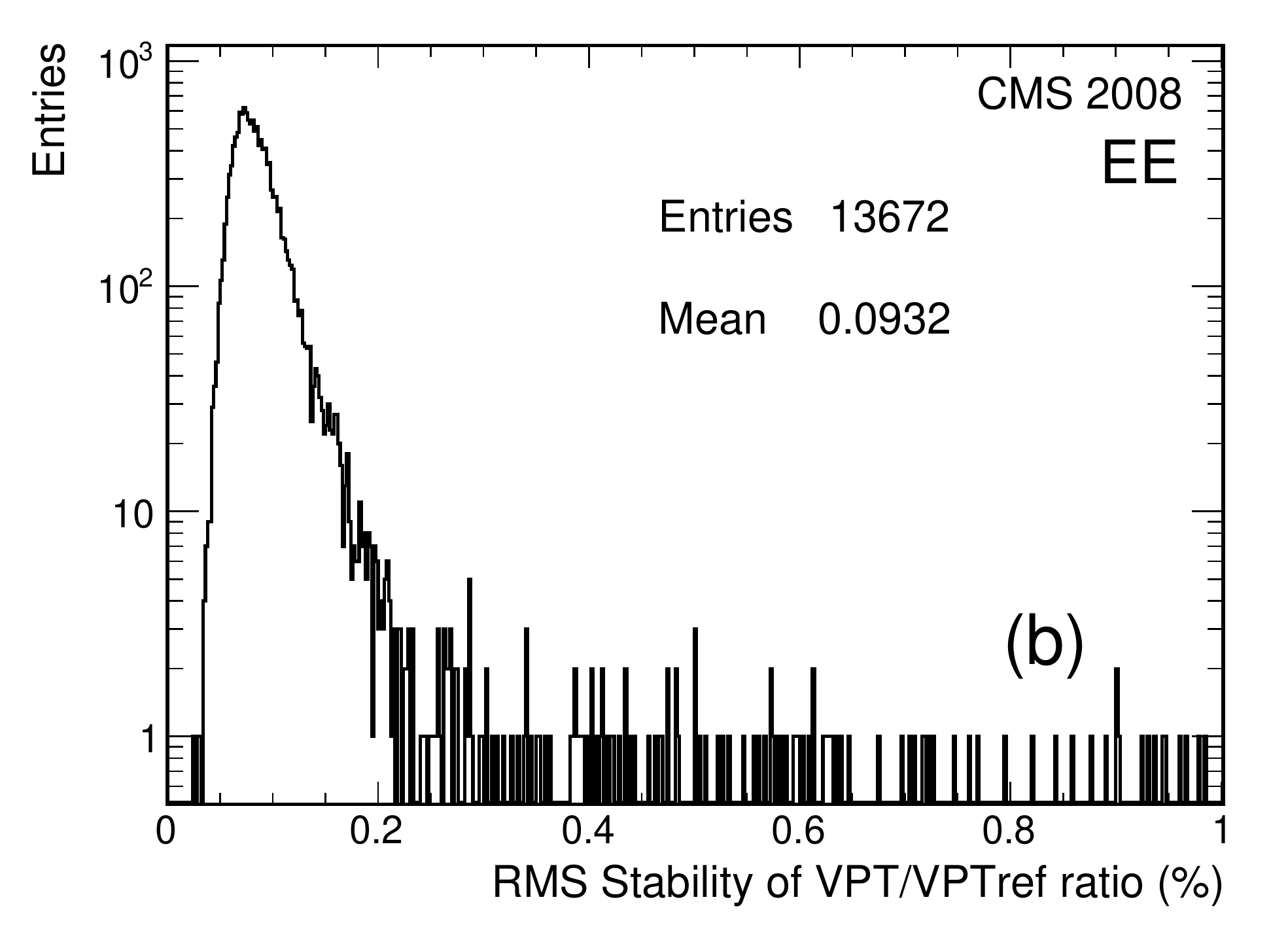}
    \hspace{1cm}
    \caption{Stability of the ECAL laser monitoring system during CRAFT. (a) RMS deviation of the quantity $\langle$APD/APD$_{\mathrm{ref}}\rangle$ for  EB channels with nominal data quality cuts applied. The most stable of the two reference APDs was used in each light monitoring region. The small secondary peak at 0.2\% from six neighbouring trigger towers (150 channels) is shown by the grey histogram. (b) Same as (a) but for the quantity $\langle$VPT/VPT$_{\mathrm{ref}}\rangle$ calculated for EE channels.}
    \label{fig:eb_laser}
  \end{center}
\end{figure}

In EB, the average over 600 events of the APD to PN response ratio, $\langle$APD/PN$\rangle$, for data taken with APD G50, was monitored, to follow variations of channel response to blue laser light. Because of problems in reading out the PN diode data during the calibration sequence (which were solved after the CRAFT run), two reference APDs were instead chosen in each light monitoring region
(approximately 200 crystals).  The reconstructed laser amplitudes in the other APDs were normalised relative to these reference channels, in order to correct for pulse-to-pulse variations in the laser output. Here it is assumed that the reference APDs are stable reference points in the data taking conditions of CRAFT, where no crystal transparency changes are expected to have happened. 
Figure~\ref{fig:eb_laser}(a) shows the RMS of the quantity $\langle$APD/APD$_{\mathrm{ref}}\rangle$  for 57\,306 channels in EB over a 200 hour long period within CRAFT, when nominal data quality conditions were met. Data from two supermodules (3400 channels) were excluded from this analysis because of low voltage supply problems during this time period. This plot illustrates the performance of the LM system: 99.8\% of the monitored channels exhibited an $\langle$APD/APD$_{\mathrm{ref}}\rangle$ stability better than the ECAL requirement of $0.2\%$. Considering all laser data recorded during CRAFT over the entire 700 hour period, $98\%$ of the monitored channels satisfied this requirement. 

The secondary peak in Fig.~\ref{fig:eb_laser}(a) arises from six neighbouring trigger towers in one light monitoring region. These trigger towers, shown by the dashed histogram in Fig.~\ref{fig:eb_laser}(a), have an average stability of 0.2\% (RMS). This is higher than most of the monitored channels in CRAFT, but remains compatible with the stability requirement. The underlying cause is a 0.6\% jump in the response of the trigger tower which provides the reference APD for this light monitoring region. The reason for this jump remains under investigation, since no corresponding fluctuation in the low voltage, high voltage or temperature readings for these channels was observed during CRAFT.  

For EE, the reconstructed laser amplitudes were normalised to a reference VPT in each supercrystal. Figure~\ref{fig:eb_laser}(b) shows the RMS of $\langle$VPT/VPT$_{\mathrm{ref}}\rangle$ over 600 events from the same data taking period, as shown in Fig.~\ref{fig:eb_laser}(a). A total of 13\,672 endcap channels were monitored. During CRAFT the average amplitude from laser light in the endcap crystals was significantly reduced from the values expected for nominal data taking (since the end of CRAFT these amplitudes have been increased by a factor of 10). As a result, approximately 1000 channels were rejected from this analysis, since their laser amplitudes during CRAFT were too low for reliable stability measurements.  In these data, 98.3\% of the monitored endcap channels showed a stability better than $0.2\%$. A significant fraction of the channels with a stability worse than 0.2\% arise from groups of five VFE cards corresponding to a single front-end card/supercrystal.  Some correlation with supercrystals which had known high voltage supply problems during CRAFT was observed, although no unique explanation for these less stable regions has been found.

\section{Validation of pre-calibration constants}\label{sec:calib}

 The channel response uniformity directly impacts on the constant term of the EM energy resolution. This uniformity depends on the accuracy of the calibration of the relative response for all channels in the detector. Inter-calibration constants are used to correct for channel-to-channel response variations, for example due to differences in crystal light yield and photodetector gain. A set of constants derived from laboratory and test beam measurements, termed pre-calibration constants, are currently used to equalise the channel-to-channel response for both the barrel and the endcaps.

 Prior to installation in the underground cavern, 9 of the 36 barrel supermodules were calibrated with 90--120 GeV electrons at the H4 test beam at CERN~\cite{1748-0221-3-10-P10007}, with an achieved precision on the relative channel-to-channel response of $0.3\%$. The remaining 27 supermodules were calibrated in the laboratory using cosmic-ray muons, with a precision of $1.5-2.5\%$~\cite{1748-0221-3-10-P10007}. For the endcap dees, the pre-calibration constants were determined from laboratory measurements of crystal light yield and VPT response. A set of 460 endcap crystals was also inter-calibrated with a precision of better than 1\% in an electron test beam during 2007. A representative subset of 162 crystals was also used to estimate the precision of the laboratory light yield and VPT response measurements. This group of crystals comes from the manufactured sample that has the best understood light yield. This sample comprises more than 80\% of the ECAL crystals.  For these 162 crystals, the combination of light yield and VPT response measurements were verified with a precision of $7.4\%$ (RMS) by comparing the laboratory and test beam measurements. 

The ultimate inter-calibration precision will be obtained from data upon LHC startup.
Data collected in 2008 from cosmic-ray muons in CRAFT and beam-induced muons during LHC operation in September were used to perform an in situ check of the pre-calibration constants obtained from laboratory measurements. The precision of these measurements, which are made at the level of 1--2\% for the barrel and better than 10\% in the endcaps, are comparable to the laboratory measurements and are therefore sufficient for LHC startup. They will also provide the initial calibration constants for the calibration methods using LHC beam events, which will ultimately achieve the final calibration goal of 0.5\%~\cite{Daskalakis:2006dx}.

\subsection{Validation of ECAL barrel pre-calibration constants}\label{sec:dedx}

A check of the pre-calibration constants for 14 of the 36 barrel supermodules was performed by comparing the stopping power (${\rm d}E/{\rm d}x$) distributions for cosmic-ray muons after the constants were applied. The sixteen supermodules located at the top and bottom of the ECAL, which have the highest acceptance to the vertical cosmic-ray muon flux, were selected for this analysis. Two supermodules were subsequently excluded due to low voltage supply problems encountered during CRAFT.  Muons with momentum between 5 and 10 GeV/$c$ were used.  In this momentum region, energy loss by ionisation is the dominant process. The muons were required to pass through the tracker volume, and only events recorded with APD G200 were used. These requirements reduce the sample from 88 million events to approximately 500\,000 events.

The momentum selection of the cosmic-ray muons is performed after the muons have passed through
the upper hemisphere but before they pass through the lower hemisphere of ECAL.
This causes a difference in the energy deposits in the two hemispheres
of about 0.5\%, due to the dependence of ${\rm d}E/{\rm d}x$ on the muon momentum.
 In order to compare the ECAL response in the upper
and lower hemispheres, this effect is corrected for in the analysis.
It was required in addition that the angle between the muon trajectory extrapolated from the tracker and the crystal axis is less than 30$^{\circ}$. This reduces systematic biases on the energy scale due to crystal energy deposits falling below the clustering or zero suppression thresholds, which is more probable for large angle tracks which pass through multiple crystals~\cite{ecaldedx}. A total of 250\,000 events remained after all selection cuts.

\begin{figure}[hbtp]
  \begin{center}
    \includegraphics[width=0.5\textwidth]{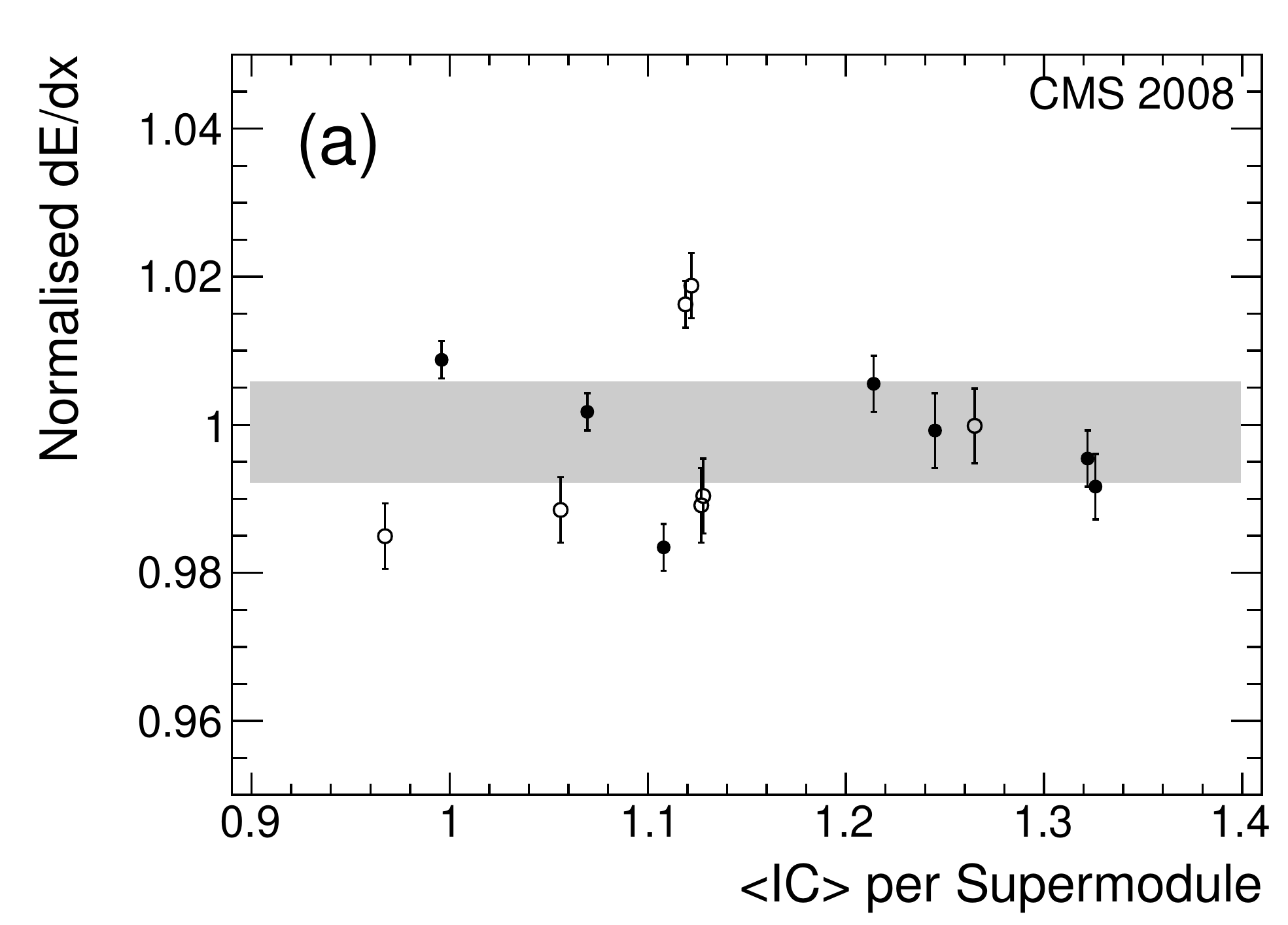}\includegraphics[width=0.5\textwidth]{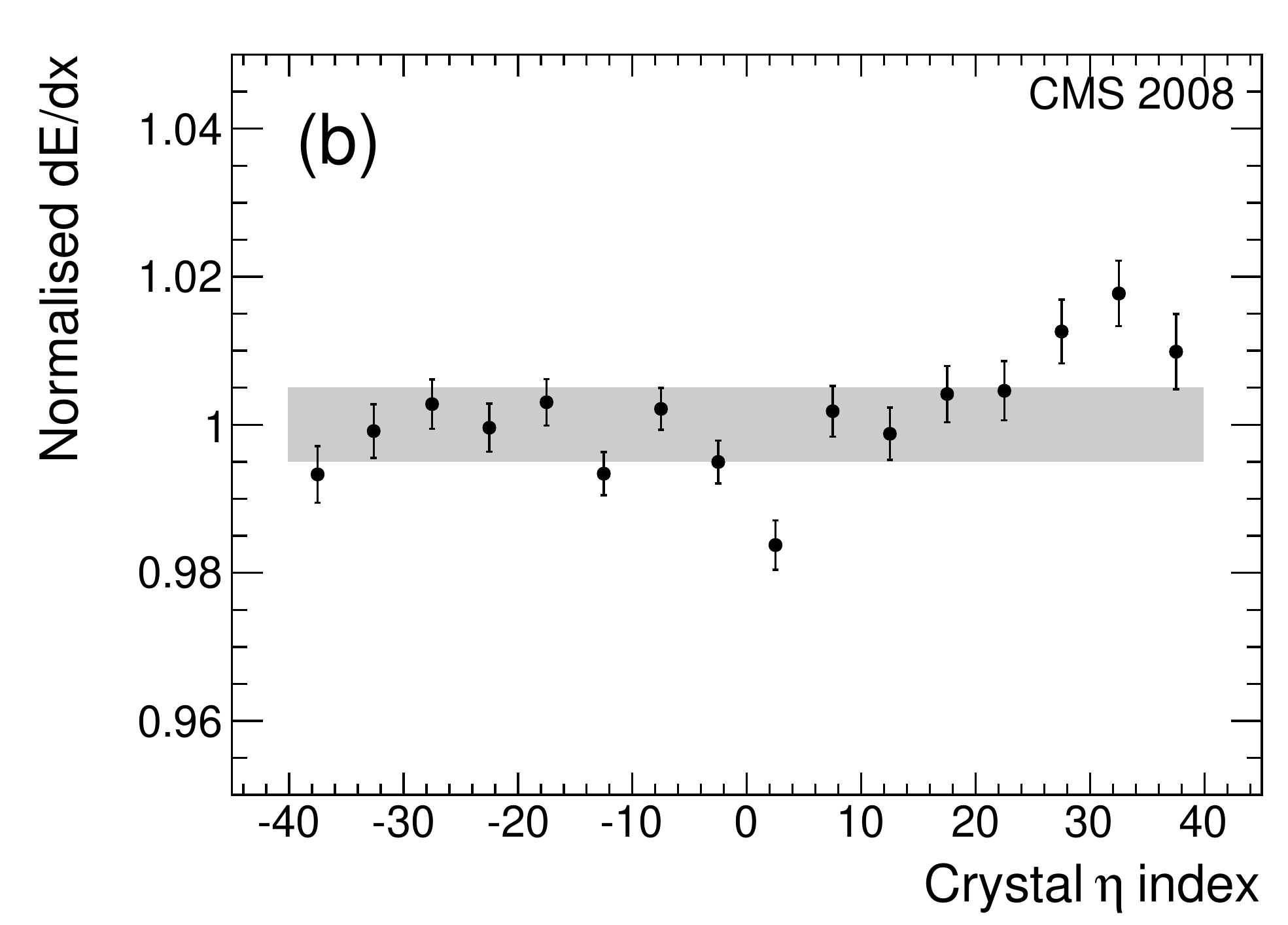}
    \caption{(a) Mean stopping power, ${\rm d}E/{\rm d}x$,  versus the mean pre-calibration constants, $\langle\mathrm{IC}\rangle$, for 14 supermodules. Each point is normalised to the average value of ${\rm d}E/{\rm d}x$ calculated using all 14 supermodules. The filled circles indicate supermodules located in the upper hemisphere of the ECAL and the open circles represent supermodules located in the lower hemisphere. (b) Mean stopping power, ${\rm d}E/{\rm d}x$, versus the crystal index in the $\eta$ coordinate. Each data point is integrated over five crystals in $\eta$ and all values of $\phi$. In both plots, the shaded region represents the systematic uncertainty on the measurement of ${\rm d}E/{\rm d}x$.}
    \label{fig:eb_calib}
  \end{center}
\end{figure}

The average pre-calibration constants for each supermodule, $\langle\mathrm{IC}\rangle$, vary by up to 30\%, due to differences in crystal light yield.  The measured ${\rm d}E/{\rm d}x$ distributions for the 14 supermodules were compared after applying the pre-calibration constants to equalise the light yield response. Figure~\ref{fig:eb_calib}(a) shows the mean stopping power for each supermodule, plotted as a function of $\langle\mathrm{IC}\rangle$. Each point is normalised to the average ${\rm d}E/{\rm d}x$ value for all 14 supermodules, and the values of $\langle\mathrm{IC}\rangle$ are normalised to a reference supermodule. The most probable value of ${\rm d}E/{\rm d}x$ in this momentum range is measured to be approximately 1.75~MeV g$^{-1}$cm$^{2}$~\cite{ecaldedx}. This corresponds to an energy loss of 335~MeV for a particle traversing the full length of a crystal.  A truncated mean is used in the determination of the average ${\rm d}E/{\rm d}x$ value in order to remove statistical fluctuations from high energy deposits in the upper 5\% of the ${\rm d}E/{\rm d}x$ distributions. The spread of these measurements, which
indicates the level of uniformity of the detector response, is about 1.1\% (RMS). This is comparable to the statistical precision of the measurements (typically 0.4\%) combined with the following systematic uncertainties: a) the dependence of the muon energy scale on the angle between the crystal axis and the muon direction (estimated to be 0.5\%); b) the variation in average muon momentum for different supermodules, since they have different angular acceptance to cosmic-ray muons and hence sample different regions of the cosmic-ray muon flux (estimated to be 0.4\%). The total systematic uncertainty of 0.6\% is indicated by the shaded band in Fig.~\ref{fig:eb_calib}(a). All estimates of systematic error are derived from data. A full description of their evaluation is provided in Ref.~\cite{ecaldedx}.

The calibration procedures in $\phi$ that utilise LHC data will yield precise inter-calibration of crystals at a given $\eta$ value. The pre-calibration constants will provide the relative scale for crystals at different $\eta$ values at LHC startup. The cosmic-ray muon data taken during CRAFT were therefore used to validate in situ the pre-calibration constants as a function of $\eta$. Figure~\ref{fig:eb_calib}(b) shows the (truncated) mean ${\rm d}E/{\rm d}x$ as a function of the crystal index in the $\eta$ coordinate. These measurements are normalised to the average ${\rm d}E/{\rm d}x$ integrated over all $\eta$ values. The distribution is plotted over the range $-0.7<\eta<0.7$, where most of the muons that pass through both the tracker and the ECAL are located.  The spread of the measurements, indicating the precision to which the $\eta$-dependent pre-calibration scale is verified, is $0.8\%$ (RMS). The statistical precision of the measurements, indicated by the error bars on the points, is typically $0.4$\%.  The total systematic uncertainty, which is represented by the shaded region, is 0.5\%. The main contribution to the systematic error is the energy scale dependence on the angle between the muon trajectory and the crystal axis (0.5\%). Since each data point integrates over all values of $\phi$, the systematic uncertainty on the muon momentum scale due to the variation in acceptance to the cosmic-ray muon flux is reduced, and is estimated to be 0.1\% in Fig.~\ref{fig:eb_calib}(b).  

\subsection{Validation of ECAL endcap pre-calibration constants}

A check of the endcap pre-calibration constants was performed using beam-induced muons, from 41 events recorded by CMS without magnetic field during LHC beam commissioning, in September 2008. The spray of  $\mathcal{O}(10^{5})$ muons produced from the LHC primary beams impinging on collimator blocks upstream of the CMS detector produced large (TeV) energy deposits in EB and EE, illuminating all active channels. In EE, the average energy per crystal was approximately 5 GeV/event. From this energy deposition, a set of local calibration coefficients was first defined, which equalise the response over a $5\times5$ crystal matrix (supercrystal),

\begin{equation}
c_{i,{\rm local}}=\frac{\langle E_{i}\rangle_{5\times 5}}{E_{i}}\quad ,   
\end{equation}\label{eqn:splash1}

where $E_{i}$ is the energy deposited in a single crystal, and $\langle E_{i}\rangle_{5\times 5}$ is the average energy recorded in the supercrystal. Here, it is explicitly assumed that the energy deposition is uniform over each supercrystal region, which is supported by the observed spatial distribution of energy deposits recorded in the endcaps.  

Inter-calibration between supercrystals was provided by the pre-calibration constants, which account for the radial dependence of the calibration coefficients due to the known variation in VPT response across the endcaps, 

\begin{equation}
c_{i}=c_{i,{\rm local}}\frac{\langle c_{i,{\rm pre}}\rangle_{5\times 5}}{\langle c_{i,{\rm local}}\rangle_{5\times 5}}\quad ,   
\end{equation}\label{eqn:splash2}

where $\langle c_{i,{\rm local}}\rangle_{5\times 5}$ and $\langle c_{i,{\rm pre}}\rangle_{5\times 5}$ are the calibration coefficients, averaged over a $5\times5$ crystal region, from beam-induced muons and laboratory measurements, respectively. 

\begin{figure}[hbtp]
  \begin{center}
    \includegraphics[width=0.33\textwidth]{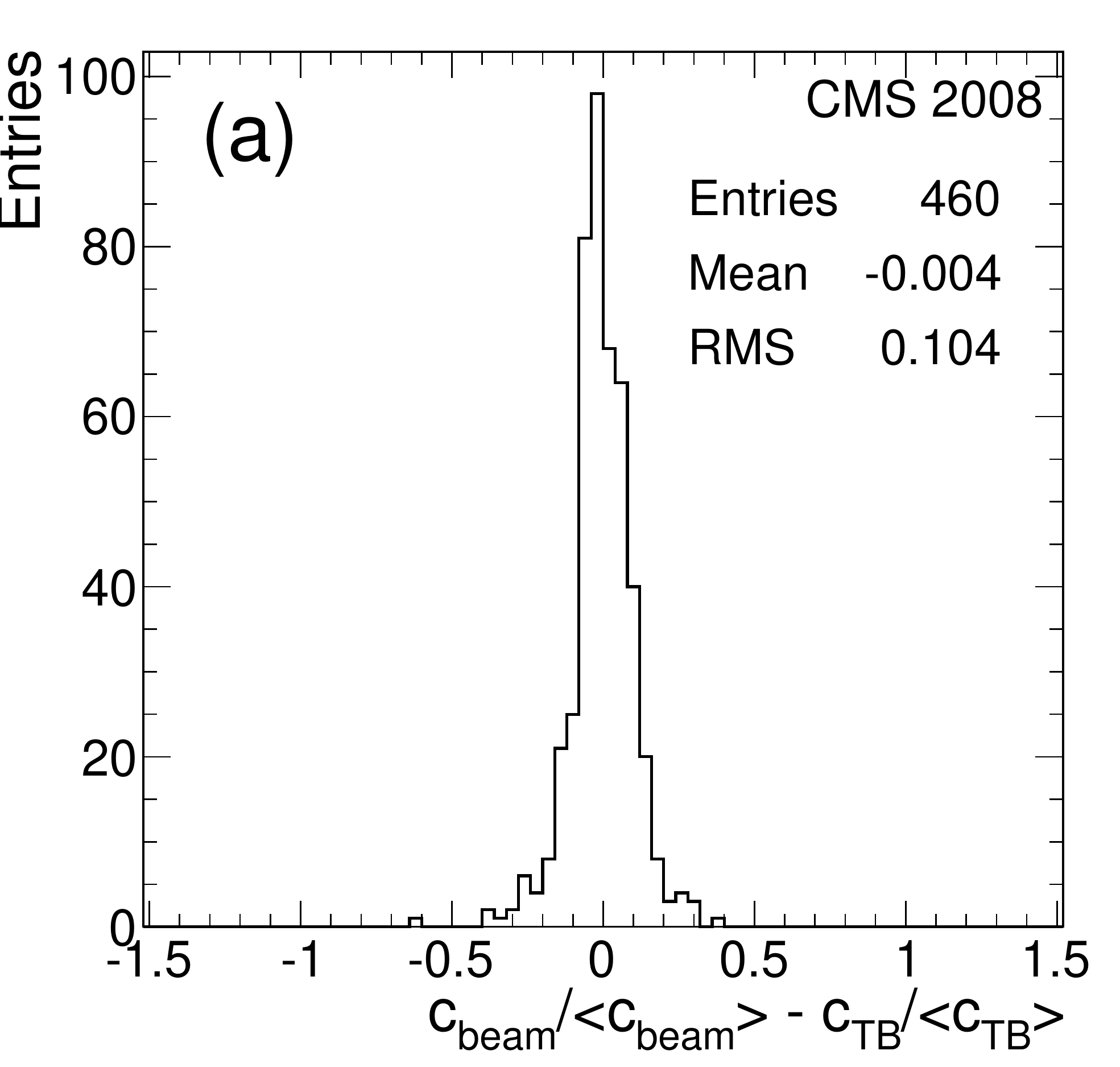}\includegraphics[width=0.33\textwidth]{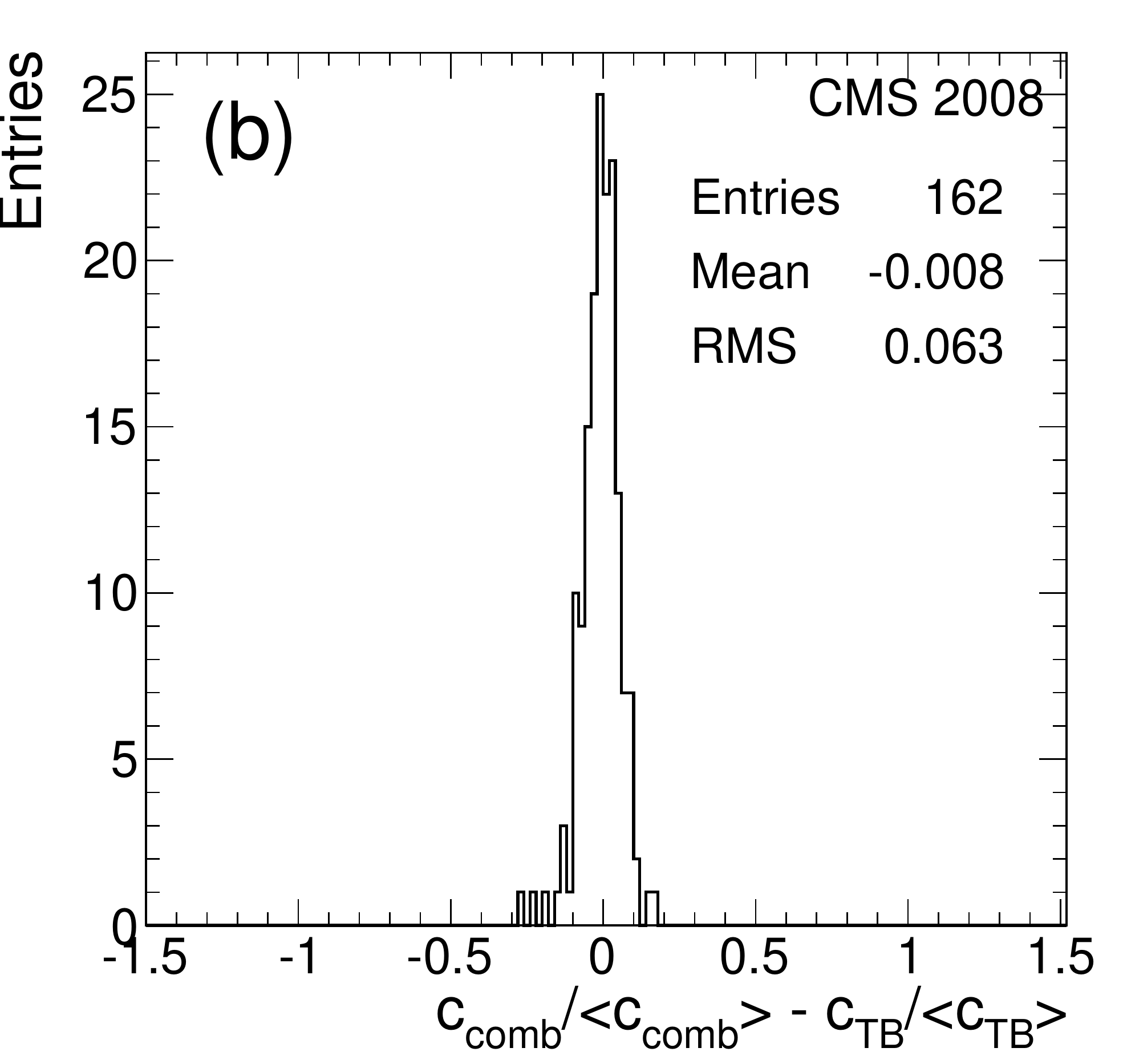}\includegraphics[width=0.33\textwidth]{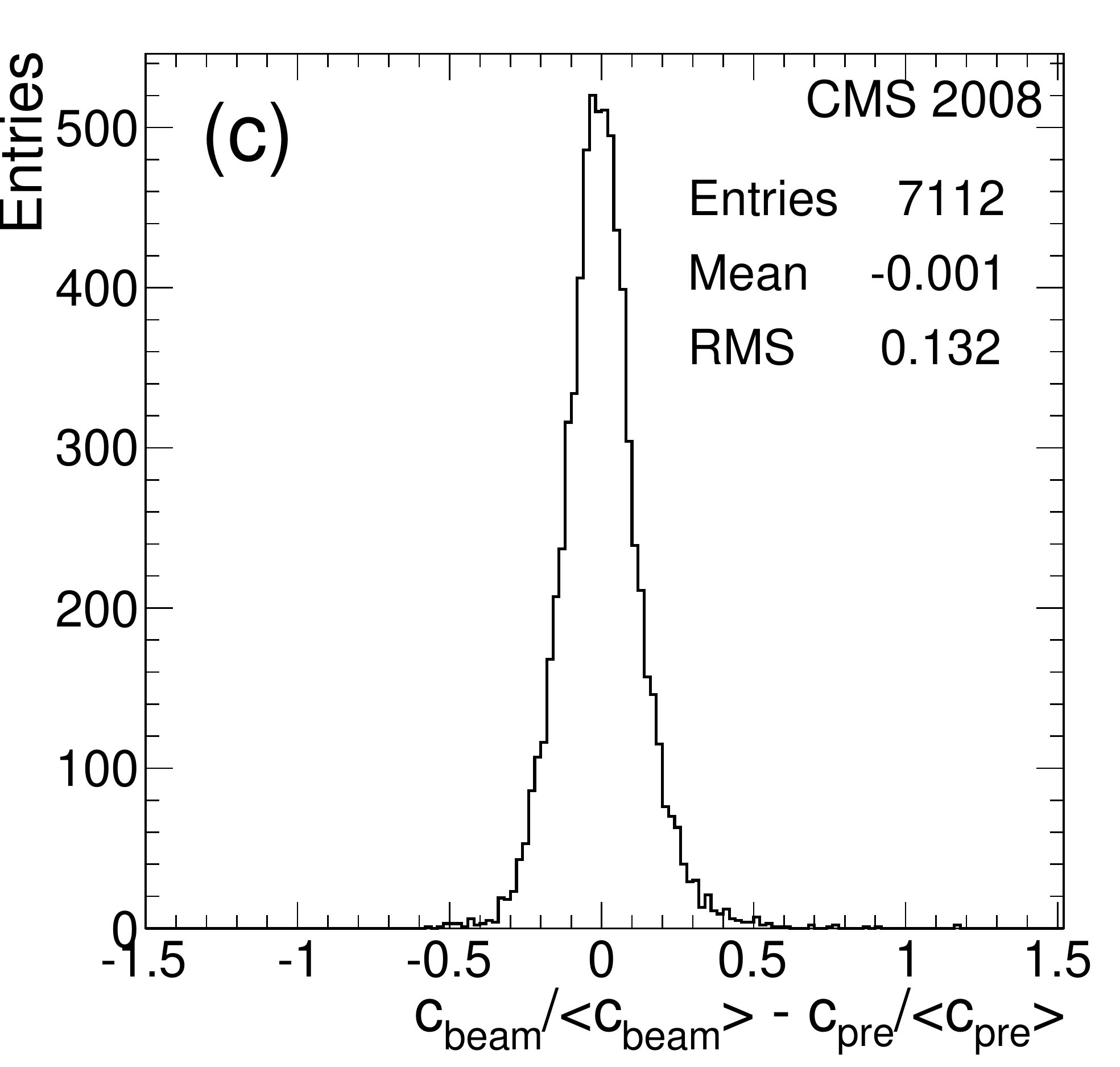}
   
    \caption{Validation of EE pre-calibration constants using beam-induced muons. (a) Comparison between normalised beam-induced muon and test beam coefficients for 460 crystals. (b) Comparison of the normalised combined beam-induced muon and pre-calibration coefficients to those derived from test beam data for the reference sample of 162 crystals. (c) Comparison between normalised beam-induced muon and pre-calibration constants for one endcap.}
    \label{fig:ee_calib}
  \end{center}
\end{figure}

Figure~\ref{fig:ee_calib}(a) compares the inter-calibration constants obtained using this method to those obtained from test beam measurements of 460 endcap crystals. The difference between the coefficients, normalised to the average value for the full sample, for the two sets of measurements, was computed for each crystal. The agreement is within 10.4\% (RMS). The statistical and systematic precision of the constants derived from beam-induced muons was investigated. The precision of these constants was evaluated with respect to the test beam measurements for an independent set of $N$ events using beam-induced muons entering from either side of the detector. The precision to which the constants were determined as a function of $N$ independent events was found to require a constant term of 8.8\% in addition to the expected $1/\sqrt{N}$ dependence. This constant term is believed to be due to non-uniformity of the energy deposition by the beam-induced muon events.

The weighted average of the pre-calibration and beam-induced muon coefficients was computed for all crystals. This weighted average is compared in
Figure~\ref{fig:ee_calib}(b) to the calibration constants obtained from the  test beam, for the reference sample of 162 crystals.  An improvement in the RMS from 7.4\% to 6.3\% was observed after combining the coefficients. This indicates that the coefficients obtained from laboratory and beam-induced muon measurements are largely independent. Figure~\ref{fig:ee_calib}(c) shows the comparison of beam-induced muon and pre-calibration constants for 7112 crystals in one endcap. Approximately 100 channels were excluded from this plot due to signal timing problems during the beam muon runs, or due to high pedestal noise. The RMS of 13.2\% is consistent with the sum in quadrature of the 10.4\% uncertainty on the beam-induced muon measurements (Fig.~\ref{fig:ee_calib}(a)) and the 7.4\% uncertainty on the pre-calibration measurements. A similar level of agreement was also observed in the other endcap. With this measurement, it is possible to deduce that the 6.3\% precision of the combined beam-induced muon and pre-calibration coefficients over the 162 reference crystals that were exposed to the test beam, is valid over all endcap crystals.

\section{Vacuum phototriode performance at 3.8~T}\label{sec:six}

Laboratory measurements of VPT performance have shown that these devices are able to operate in a high magnetic field environment, such as present in CMS~\cite{Bell:2004eu}. Measurements taken during 2008 with the ECAL laser and LED monitoring systems in the CMS underground cavern, in an operating field of 3.8~T, were used to study the performance of the 14\,648 installed VPTs. They confirmed the operability of these devices in a high field and permitted studies of the effects of magnetic field and pulsing rate on the VPT anode sensitivity.

\subsection{VPT response as a function of orientation to the magnetic field direction}

Over the range of angles between the endcap VPT tube axes and the magnetic field direction (4 to 18 degrees), the VPT anode sensitivity changes by a value between 5 and 30\%, relative to the response at 0~T. In order to measure this effect, a series of laser runs were taken in both endcaps at zero and 3.8~T magnetic field. Since the pre-calibration constants for the endcaps discussed in the previous section and the energy scale derived from test beam measurements were all obtained at zero magnetic field, the laser data were used to translate the pre-calibration constants to 3.8~T. 

\begin{figure}[hbtp]
  \begin{center}
    \includegraphics[width=0.8\textwidth]{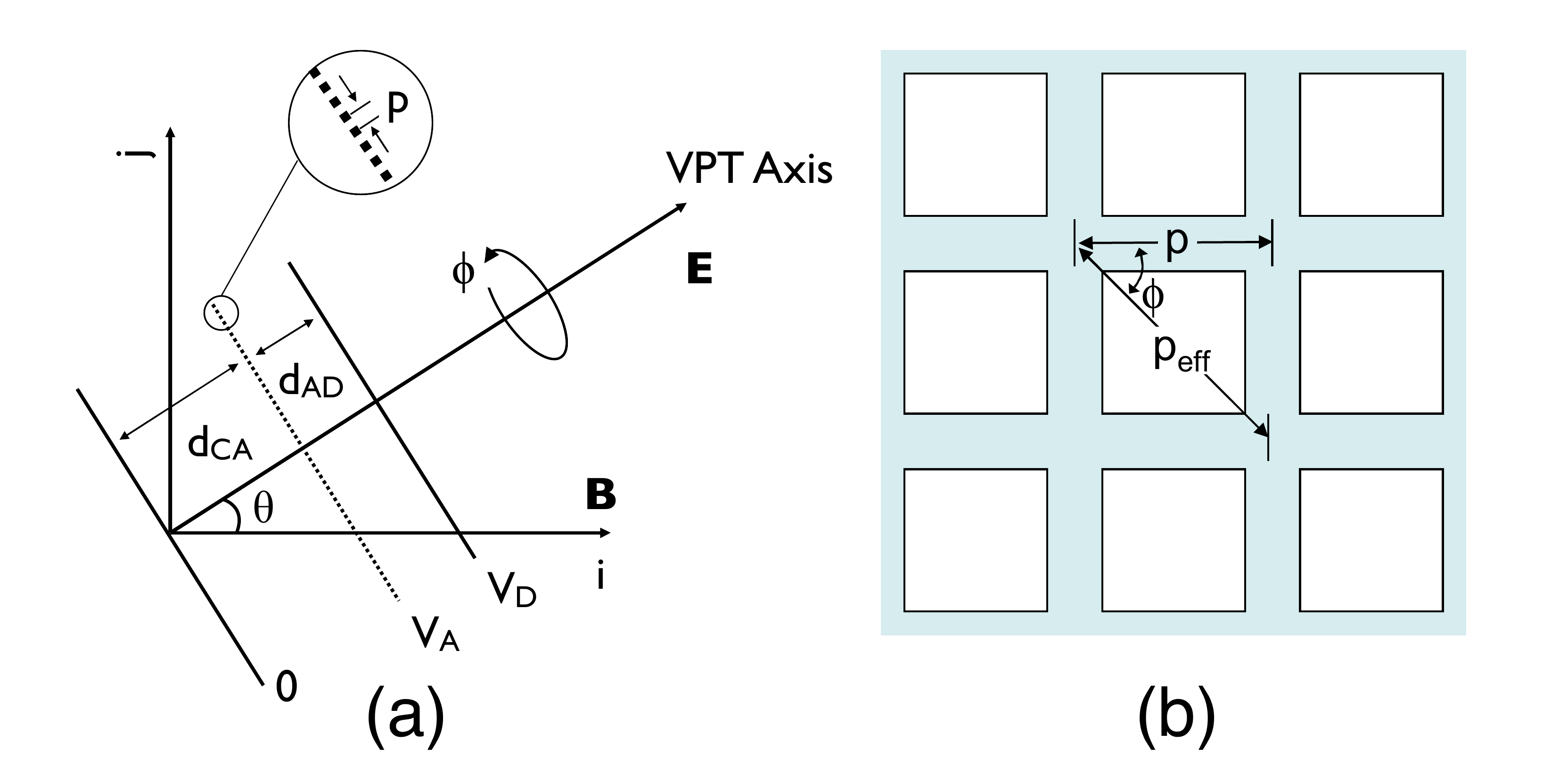}
    \caption{(a) Schematic representation of the electrode structure of a VPT, with the device axis, $\vec{E}$, at an angle $\theta$ to the magnetic field direction,  $\vec{B}$, which is here assumed to be parallel to the axis labelled $i$.  The anode grid and dynode are maintained at potentials $V_{A}$ and $V_{D}$, respectively. (b) Schematic representation of the anode grid, showing how the effective pitch, $p_{\rm eff}$, varies as the VPT is rotated through an angle $\phi$ about its axis.}
    \label{fig:vpt}
  \end{center}
\end{figure}

\begin{figure}[hbtp]
  \begin{center}
    \includegraphics[width=0.5\textwidth]{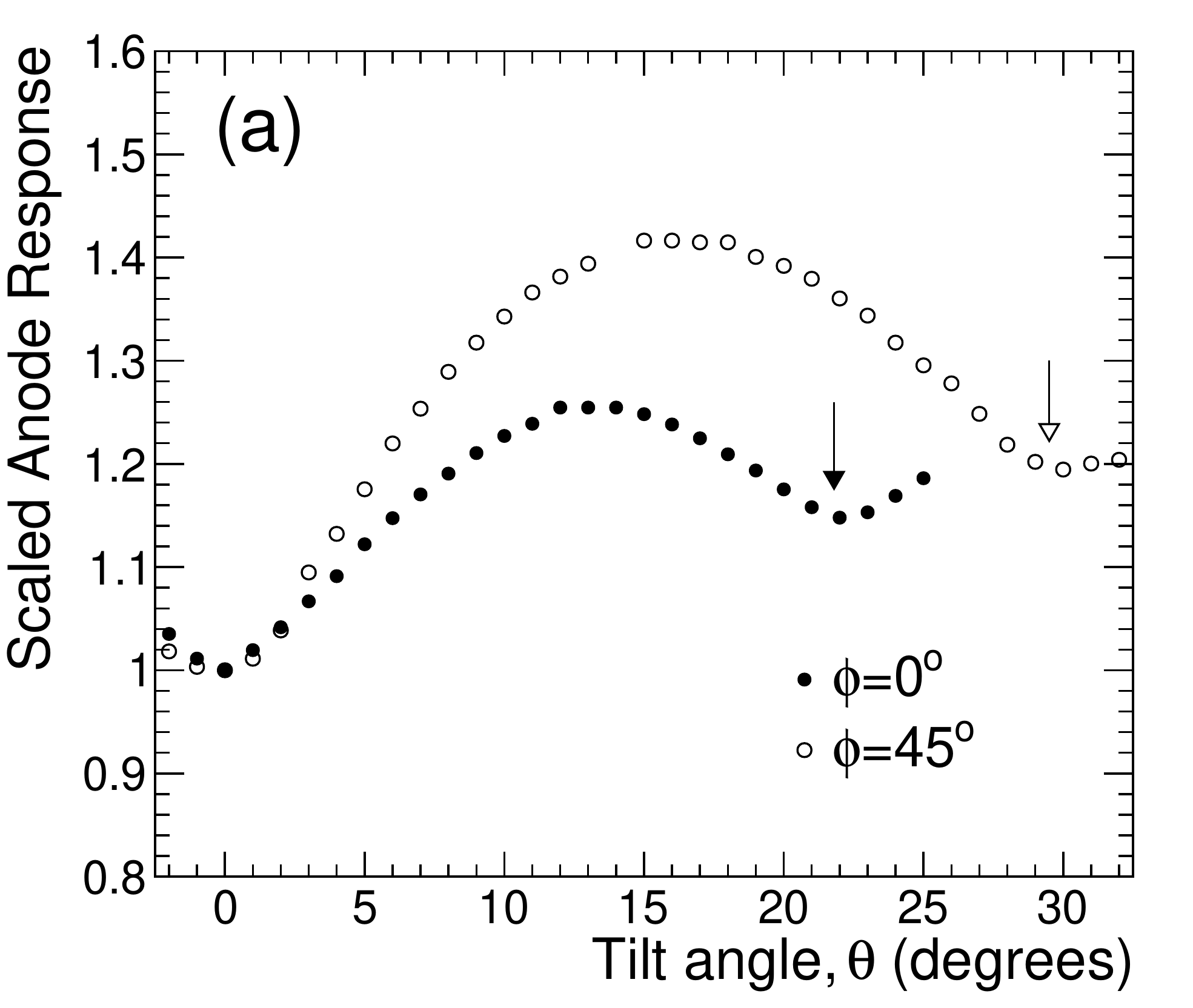}\includegraphics[width=0.5\textwidth]{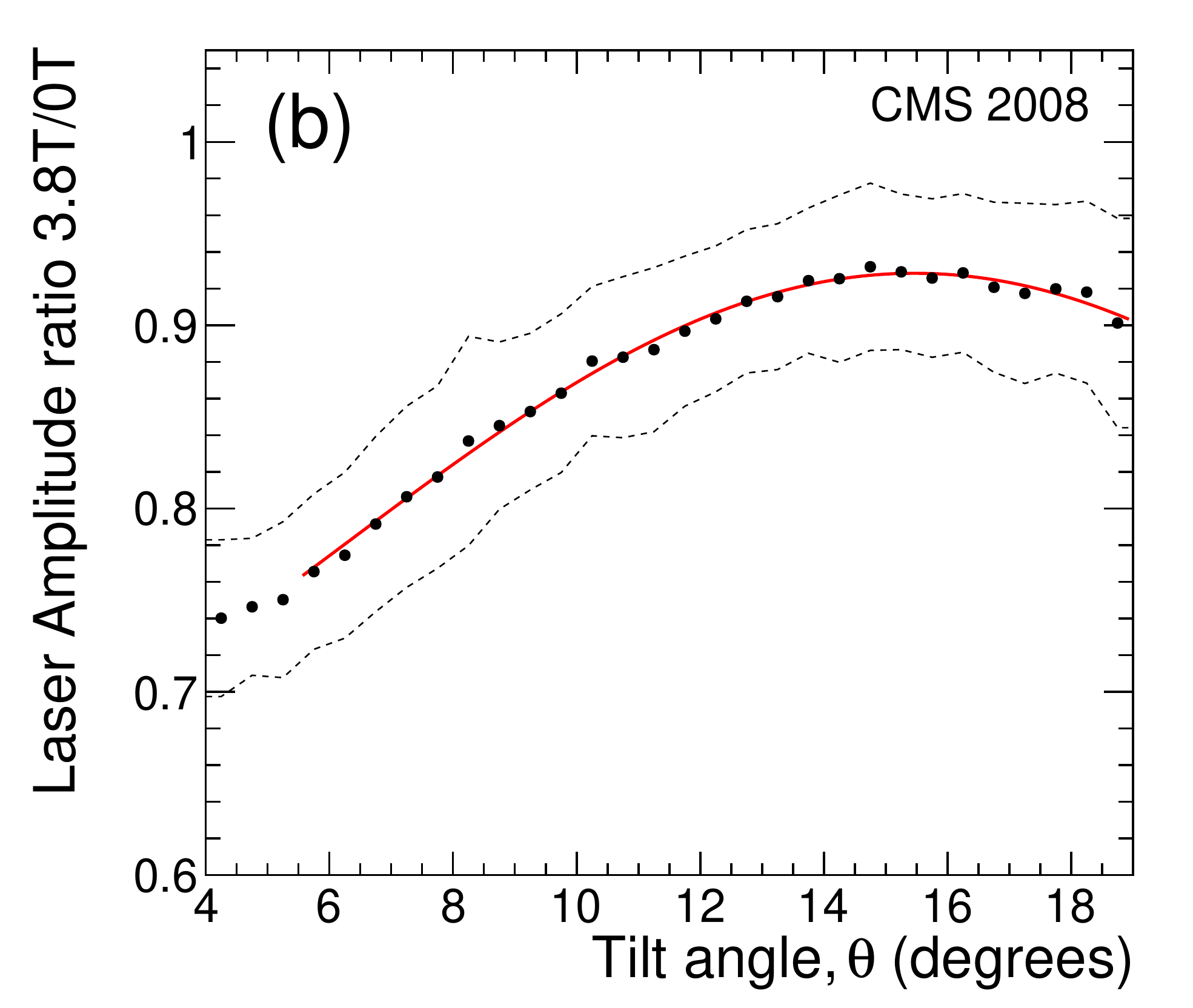}
     
    \caption{(a) Laboratory measurement of the response of a single VPT as a function of the tilt angle $\theta$ in a 3.8~T axial magnetic field, for two orientation angles, $\phi$, of the VPT grid relative to the magnetic field axis. Both sets of data were normalised to unity at $\theta=0$. The arrows mark the predicted positions of the minima in the VPT response as a function of $\theta$ for the two $\phi$ orientations, using the model described in Ref.~\cite{CMS_NOTE_2009-014}. (b) Normalised ratio of VPT response  $Y_{3.8}/Y_0$ as a function of the tilt angle $\theta$ for EE laser runs taken with a magnetic field of 3.8~T and 0~T. The band between the two dashed lines represents the RMS spread of the quantity  $Y_{3.8}/Y_0$ for all VPTs at a given $\theta$ angle. The solid line shows the result of a fit to the data using Eq.~(\ref{eqn:vptfit}).}
    \label{fig:ee_vpt}
  \end{center}
\end{figure}

A schematic representation of the disposition of the electrode structure of a VPT in a magnetic field is shown in Fig.~\ref{fig:vpt}(a). The response varies as a function of the angle $\theta$ between the axis of the device and the magnetic field direction, and as a function of the orientation $\phi$ of the device about its axis. In general, the response curve exhibits two main features: a plateau, modulated by a series of minima centred on $\theta=0$, and a sharp fall-off at larger values of $|\theta|$~\cite{CMS_NOTE_2009-014}.  Both of these features depend on the physical structure (pitch and thickness) of the anode grid (see Fig.~\ref{fig:vpt}(b)). Only the plateau is relevant to the operation in CMS, since the ultimate fall-off occurs outside of the range of angles encountered in the ECAL endcaps. Figure~\ref{fig:ee_vpt}(a) shows the normalised anode response as a function of $\theta$, measured in the laboratory at 3.8~T, for a tube with anode grid orientations $\phi=0$ or $45^{\circ}$. Here, local minima in the response curves are clearly seen. The minima are also shown to depend on the rotation angle, $\phi$, of the grid. Since $\phi$ was randomised during dee construction, this will result in the smearing of the distribution of VPT response at 3.8~T for tubes at a fixed value of $\theta$.

 The dip/peak structure results from secondary electrons drifting in the direction defined by $\vec{E}\times \vec{B}$ (perpendicular to the plane of the paper in Fig.~\ref{fig:vpt}(a)). An analytical model has been developed from this concept that enables the position of the minima to be predicted~\cite{CMS_NOTE_2009-014}. For the VPTs operating at 3.8~T, with a difference in anode-dynode potential of 200~V, the first dip is predicted to occur at  an angle $\tan(\theta_{1})=\tan(21.8^{\circ})/\cos(\phi)$. The $\phi$-dependence results from the change in the effective anode grid pitch as a function of the $\phi$ rotation angle, as shown in Fig.~\ref{fig:vpt}(b). The predicted dip positions for the two angle scans shown in Fig.~\ref{fig:ee_vpt}(a), which are represented by the two arrows, show good agreement with the laboratory data.

  Since PN diode readout was not available during CRAFT, the endcap laser amplitudes used for this analysis were normalised using the laser amplitude measurements from the barrel supermodules. Since the barrel measurements are stable with respect to the magnetic field, such a normalisation suppresses amplitude variations due to the
laser light source while preserving the variations of the endcap laser
amplitude due to the magnetic field.

 The measured dependence of VPT response as a function of the tilt angle $\theta$ of the endcap VPTs with respect to the magnetic field direction is shown in Fig.~\ref{fig:ee_vpt}(b). The ratio of VPT response for two laser runs taken at 3.8~T ($Y_{3.8}$) and 0~T ($Y_0$) during CRAFT is shown, for the angular range between 4 and 18 degrees. Among the endcap VPTs, more than 75\% of the tubes exhibit tilt angles between 10 and 18 degrees, and the measured value of the ratio $Y_{3.8}/Y_0$ averaged over all tubes is 88.9\%. The RMS spread of the data points, indicated by the dashed lines in Fig.~\ref{fig:ee_vpt}(b), shows the effect of averaging over the $\phi$ angle. 

 A fit was performed to the measured ratio $Y_{3.8}/Y_0$ using an empirically derived function of the form

\begin{equation}
f(\theta)=
 S \left[
    1 - \frac{x}{2}
    + \frac{x}{2} \sin \left(
      \frac{\theta-\theta_0}{\theta_p}
    \right)
  \right]\quad .\label{eqn:vptfit}
\end{equation}

The parameters $S$ and $x$ control the amplitude and vertical offset of the sinusoidal component of the  function, and $\theta_0$ and $\theta_p$ control the horizontal offset and period.

Individual correction factors were obtained for all tubes at a given angle $\theta$.   The precision of the measurement was estimated by examining the
dependence on the tilt angle $\theta$ of the fit to the VPT yield ratio $Y_{3.8}/Y_{0}$
 .  In addition, the stability of the correction factors was measured by applying them to other laser runs taken during CRAFT at 0~T and 3.8~T.  The estimated precision was found to be $\approx 4$\%, and is mainly due to the averaging over channels with random $\phi$ angles at a constant $\theta$ value, consistent with the spread of values indicated by the dashed lines in Fig.~\ref{fig:ee_vpt}(b). Applying these factors to the pre-calibration constants obtained at 0~T provides an 11\% average correction with a 4\% uncertainty. 

The precision of this measurement will be significantly improved in the future, when laser and LED data normalised using the PN diode readout is used to obtain per-channel normalisation factors. These will take into account both the $\theta$ and $\phi$ dependence of the VPT response in the strong magnetic field of CMS, and will eliminate the need to provide an average correction for each 
value of $\theta$. This was necessitated by the use of the ECAL barrel to normalise the laser amplitudes in the endcaps, which only provides an overall scale for the laser output, rather than a channel-by-channel normalisation. It is expected that the
data normalised by PN diode readout should provide corrections to the VPT response measured at 0~T to the CMS operating field of 3.8~T with a precision of $\approx 0.1\%$. 

Studies were also performed over a limited angular range by LED measurements taken at 0~T and 3.8~T for a single diffusing sphere (200 channels). The measured ratios $Y_{3.8}/Y_0$ for LED and laser data for these channels agree at the 2\% level, which is within the uncertainties quoted above for the laser measurements.

\subsection{VPT rate stability}\label{sec:vptrate}

The VPTs used in CMS are designed to operate in a high magnetic field. Since they do not have electrostatic focussing, they require the presence of a strong quasi-axial magnetic field for stable operation. Variations of 5 to 20\% in VPT response at zero magnetic field, induced by sudden changes in the illuminating light pulse rate, have been observed in both laboratory and test beam measurements. These variations were found to be strongly suppressed in the laboratory at 1.8~T and 4~T, and also suppressed in the presence of a constant background illumination. The LED pulser system installed in CMS can provide a constant background rate to each VPT, in order to keep them active in the absence of LHC collisions and reduce their rate sensitivity.

Tests of the VPT rate stabi§lity were carried out on 200 VPTs in CMS at 0~T and 3.8~T during late 2008. The tests were initiated with the VPTs in a quiescent state (no pulsing for the previous 12 hours). High rate LED pulsing was then turned on, delivering to individual VPTs an energy equivalent amplitude of 10--15~GeV with a 10~kHz rate. This is roughly equivalent to the expected average VPT load during LHC running at a luminosity of $10^{33}$~cm$^{-2}$s$^{-1}$. High rate pulsing continued for 17 hours and was then turned off. The response of each VPT was continuously monitored throughout the entire period via dedicated LED runs (approximately 500 pulses taken at $100$~Hz), including several hours before and after the high rate LED illumination. The LED monitoring light was simultaneously measured by the PN diodes, which were used to provide pulse-to-pulse normalisation of the signals measured by the VPTs.

\begin{figure}[hbtp]
  \begin{center}
    \includegraphics[width=0.6\textwidth]{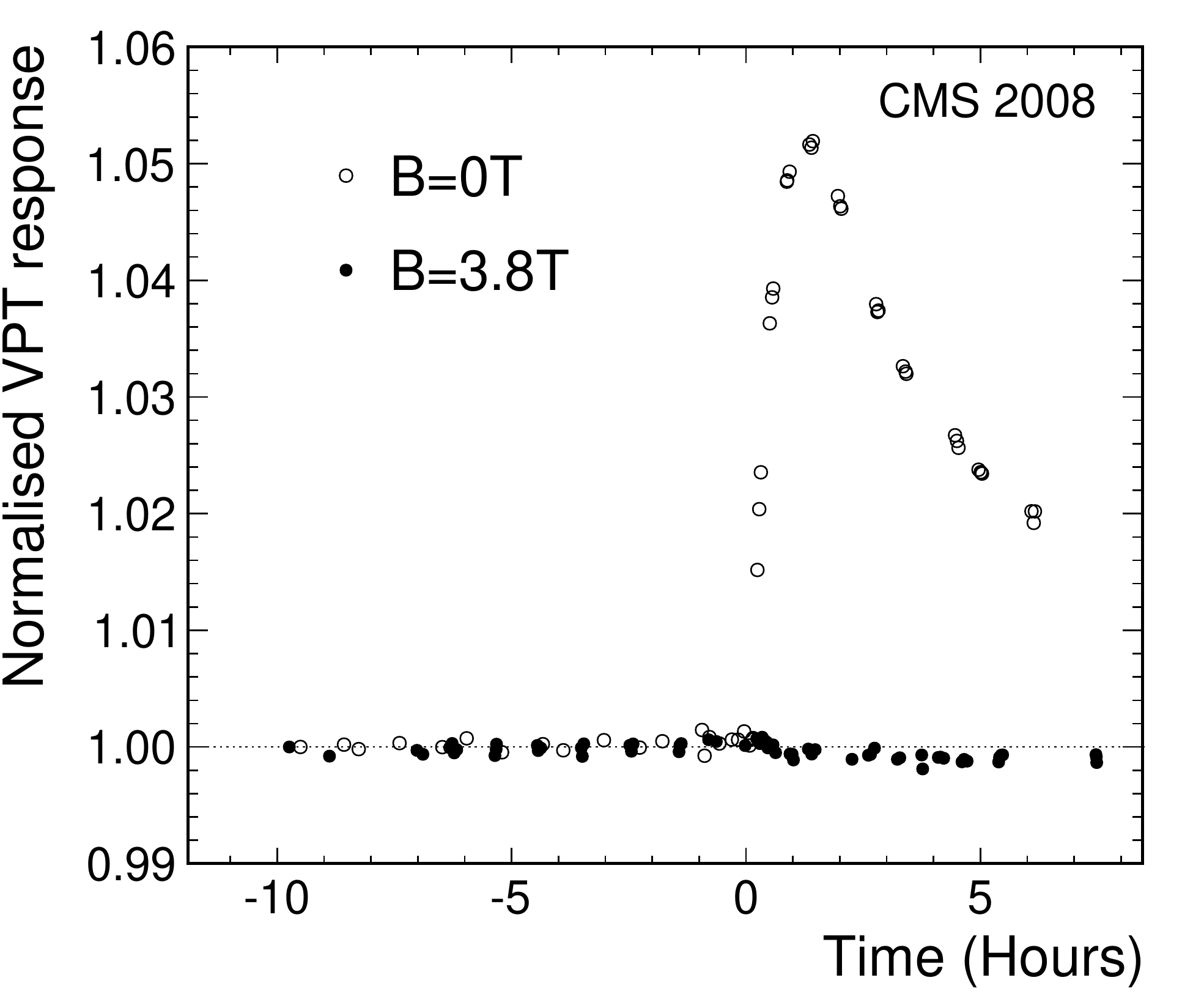}
     \hspace{1cm}
    \caption{Normalised VPT response (averaged over 200 tubes) for two high rate LED pulsing tests at 0~T (open circles) and 3.8~T (filled circles) during CRAFT. In both tests, LED pulsing with a rate of 10~kHz was performed for a period of 17 hours and turned off at time $T=0$~hours. The VPT response was normalised to the value at time $T=-10$~hours in both tests.}
    \label{fig:vpt_rate}
  \end{center}
\end{figure}

 Figure~\ref{fig:vpt_rate} shows the normalised VPT response, averaged over 200 channels, as a function of time, during two tests, performed with a magnetic field of 0~T and 3.8~T. In both cases, the LED pulsing at high rate was turned off at time $T=0$~hours. The average variation of VPT response when the high rate pulsing was turned off was measured to be 5\% with the CMS magnet at 0~T. When the same test was performed with the CMS magnet at 3.8~T, the average variation of VPT response was measured to be less than 0.2\%, as expected. During CMS operation the LED and laser light monitoring systems will be used to continuously monitor the rate sensitivity of VPTs. Further dedicated tests are planned for 2009, prior to the start of LHC operation, on a larger set of VPTs and also studying the effect on rate sensitivity of exposing the tubes to a constant level of background illumination from the LED system (at approximately 100~Hz).

%%%%%%%%%%%%%%%%%%%%%%%%%%%%%%%%%%%%%%%%%%%%%%%%%%%%%%%%%%%%%%%%%%%%%%%%%%%%%

%%%%%%%%%%%%%%%%%%%%%%%%%%%%%%%%%%%%%%%%%%%%%%%%%%%%%%%%%%%%%%%%%%%%%%%%%%%%%

\section{Summary}

The installation of the crystal ECAL in CMS was completed in August 2008 with the insertion of the two endcap detectors. The cosmic-ray data taking period in October and November 2008 was the first opportunity to operate the ECAL for an extended period of time, with CMS in its final configuration. Both the barrel and endcap detectors operated stably during this period, with more than 98.5\% of channels active. The stability of electronic noise, high voltage and temperature are found to satisfy the ECAL performance targets and therefore do not significantly contribute to the constant term of the EM energy resolution.

The ECAL calibration sequence records laser-induced events, pedestal events and test pulse data during the LHC abort gap. This was exercised for the first time in CMS during this period. The ultimate purpose of these data is to track changes in crystal transparency under irradiation with an accuracy of $0.2\%$. The data taken during the cosmic-ray tests were used to evaluate the stability of the light monitoring system in a 200 hour period only using the channels for which nominal data quality criteria were met (94\% of barrel and 93\% of endcap channels). A total of 99.8\% of the monitored barrel channels and $98.3\%$ of the monitored endcap channels showed a normalised laser amplitude stability better than 0.2\% (RMS). 

Cosmic-ray muon events and beam-induced muons in the ECAL were used to verify the pre-calibration constants in the barrel and endcaps, which were derived from laboratory and test beam measurements made prior to the installation of the detectors in the underground cavern. These constants, which will provide initial values for the crystal calibration using LHC beam data at startup, were confirmed with a precision comparable to that obtained from the laboratory measurements. In the barrel, the relative energy scale between supermodules was verified with a precision of $\approx 1\%$. In the endcaps, the precision of the constants at zero magnetic field was improved from $7.4\%$ to $6.3\%$ combining the pre-calibration coefficients with those obtained from beam-induced muons.

This data taking period was the first opportunity to operate the ECAL endcap detectors in the 3.8~T CMS magnetic field. The 14\,648 VPT photodetectors were shown to operate stably at 3.8~T. The dependence of VPT response on the angle of the tube axes with respect to the magnetic field direction was measured in situ, and used to update the existing calibration constants that were obtained at 0~T. The endcap LED system was commissioned, and was used to measure the sensitivity of the VPT anode  (averaged over 200 tubes) to sudden changes in rate. This sensitivity was found to be less than 0.2\% in the high magnetic field of CMS.

%%%%%%%%%%%%%%%%%%%%%%%%%%%%%%%%%%%%%%%%%%%%%%%%%%%%%%%%%%%%%%%%%%%%%%%%%%%%%

\section*{Acknowledgements}

We thank the technical and administrative staff at CERN and other CMS Institutes, and acknowledge support from: FMSR (Austria); FNRS and FWO (Belgium); CNPq, CAPES, FAPERJ, and FAPESP (Brazil); MES (Bulgaria); CERN; CAS, MoST, and NSFC (China); COLCIENCIAS (Colombia); MSES (Croatia); RPF (Cyprus); Academy of Sciences and NICPB (Estonia); Academy of Finland, ME, and HIP (Finland); CEA and CNRS/IN2P3 (France); BMBF, DFG, and HGF (Germany); GSRT (Greece); OTKA and NKTH (Hungary); DAE and DST (India); IPM (Iran); SFI (Ireland); INFN (Italy); NRF (Korea); LAS (Lithuania); CINVESTAV, CONACYT, SEP, and UASLP-FAI (Mexico); PAEC (Pakistan); SCSR (Poland); FCT (Portugal); JINR (Armenia, Belarus, Georgia, Ukraine, Uzbekistan); MST and MAE (Russia); MSTDS (Serbia); MICINN and CPAN (Spain); Swiss Funding Agencies (Switzerland); NSC (Taipei); TUBITAK and TAEK (Turkey); STFC (United Kingdom); DOE and NSF (USA). Individuals have received support from the Marie-Curie IEF program (European Union); the Leventis Foundation; the A. P. Sloan Foundation; and the Alexander von Humboldt Foundation.

%%%%%%%%%%%%%%%%%%%%%%%%%%%%%%%%%%%%%%%%%%%%%%%%%%%%%%%%%%%%%%%%%%%%%%%%%%%%%

\bibliography{auto_generated}   % will be created by the tdr script.

\providecommand{\href}[2]{#2}\begingroup\raggedright\begin{thebibliography}{10}

\bibitem{:2008zzk}
{CMS Collaboration}, ``{The CMS experiment at the CERN LHC}'', {\em JINST} {\bf
  3} (2008)
S08004.
%%CITATION = JINST,0803,S08004;%%.
  \href{http://dx.doi.org/10.1088/1748-0221/3/08/S08004}{{\tt
  doi:10.1088/1748-0221/3/08/S08004}}.

\bibitem{lhc}
{L. Evans and P. Bryant (eds.)}, ``{LHC Machine}'', {\em JINST} {\bf 3} (2008)
  S08001. \href{http://dx.doi.org/10.1088/1748-0221/3/08/S08001}{{\tt
  doi:10.1088/1748-0221/3/08/S08001}}.

\bibitem{CERN_LHCC_97-033}
{CMS Collaboration}, ``The Electromagnetic Calorimeter Project: Technical
  Design Report'', {\em CERN/LHCC} {\bf
  \href{http://cdsweb.cern.ch/record/349375}{97-033}} (1997). CMS TDR 4.

\bibitem{annrev}
G.~Gratta, H.~Newman, and R.~Y. Zhu, ``Crystal Calorimeters in Particle
  Physics'', {\em Annual Review of Nuclear and Particle Science} {\bf 44}
  (1994), no.~1, 453--500.
  \href{http://dx.doi.org/10.1146/annurev.ns.44.120194.002321}{{\tt
  doi:10.1146/annurev.ns.44.120194.002321}}.

\bibitem{CRAFTGeneral}
{CMS Collaboration}, ``{Commissioning of the CMS Experiment and the Cosmic Run
  at Four Tesla}'', {CMS-CFT-09-008}. {To be submitted to JINST}.

\bibitem{CERN_LHCC_2003-055}
M.~Raymond {et~al.}, ``The MGPA Electromagnetic Readout Chip for CMS'', {\em
  CERN/LHCC} {\bf \href{http://cdsweb.cern.ch/record/712053}{2003-055}} (2003).

\bibitem{Paganini:2009zz}
P.~Paganini, ``CMS Electromagnetic Trigger commissioning and first operation
  experiences'', {\em J. Phys. Conf. Ser.} {\bf 160} (2009)
012062.
%%CITATION = 00462,160,012062;%%.
  \href{http://dx.doi.org/10.1088/1742-6596/160/1/012062}{{\tt
  doi:10.1088/1742-6596/160/1/012062}}.

\bibitem{4179147}
T.~Christiansen, ``The CMS Magnet Test and Cosmic Challenge'', {\em Nuclear
  Science Symposium Conference Record, 2006. IEEE} {\bf 2} (2006) 906--908.
  \href{http://dx.doi.org/10.1109/NSSMIC.2006.355993}{{\tt
  doi:10.1109/NSSMIC.2006.355993}}.

\bibitem{CERN_LHCC_2000-038}
{CMS Collaboration}, ``The TriDAS Project Technical Design Report, Volume I:
  The Trigger Systems'', {\em CERN/LHCC} {\bf
  \href{http://cdsweb.cern.ch/record/706847}{2000-038}} (2000). CMS TDR 6.

\bibitem{CERN_LHCC_2006-021}
{CMS Collaboration}, ``CMS Physics Technical Design Report, Volume II: Physics
  Performance'', {\em CERN/LHCC} {\bf
  \href{http://cdsweb.cern.ch/record/942733}{2006-021}} (2006). CMS TDR 8.2.

\bibitem{ecaltrig}
{CMS Collaboration}, ``{Performance of the CMS Level-1 Trigger during
  Commissioning with Cosmic Rays}'', {CMS-CFT-09-013}. {To be submitted to
  JINST}.

\bibitem{Almeida:2005aa}
N.~Almeida {et~al.}, ``The Selective Read-Out Processor for the CMS
  Electromagnetic Calorimeter'', {\em IEEE Trans. Nucl. Sci.} {\bf 52} (2005)
772--777.
%%CITATION = IETNA,52,772;%%.
  \href{http://dx.doi.org/10.1109/TNS.2005.850946}{{\tt
  doi:10.1109/TNS.2005.850946}}.

\bibitem{Adzic:2006nn}
P.~Adzic {et~al.}, ``Reconstruction of the signal amplitude of the CMS
  electromagnetic calorimeter'', {\em Eur. Phys. J.} {\bf C46S1} (2006)
23--35.
%%CITATION = EPHJA,C46S1,23;%%.
  \href{http://dx.doi.org/10.1140/epjcd/s2006-02-002-x}{{\tt
  doi:10.1140/epjcd/s2006-02-002-x}}.

\bibitem{ecaltime}
{CMS Collaboration}, ``{Time Reconstruction and Performance of the CMS
  Electromagnetic Calorimeter}'', {CMS-CFT-09-006}. {To be submitted to JINST}.

\bibitem{ecaldedx}
{CMS Collaboration}, ``{Measurement of the muon stopping power of Lead
  Tungstate}'', {CMS-CFT-09-005}. {To be submitted to JINST}.

\bibitem{Adzic:2007mi}
P.~Adzic {et~al.}, ``Energy resolution of the barrel of the CMS electromagnetic
  calorimeter'', {\em JINST} {\bf 2} (2007)
P04004.
%%CITATION = JINST,2,P04004;%%.
  \href{http://dx.doi.org/10.1088/1748-0221/2/04/P04004}{{\tt
  doi:10.1088/1748-0221/2/04/P04004}}.

\bibitem{Bell:2004eu}
K.~W. Bell {et~al.}, ``Vacuum phototriodes for the CMS electromagnetic
  calorimeter endcap'', {\em IEEE Trans. Nucl. Sci.} {\bf 51} (2004)
2284--2287.
%%CITATION = IETNA,51,2284;%%.
  \href{http://dx.doi.org/10.1109/TNS.2004.836053}{{\tt
  doi:10.1109/TNS.2004.836053}}.

\bibitem{craftdataflow}
{CMS Collaboration}, ``{CMS Data Processing Workflows During an Extended Cosmic
  Ray Run}'', {CMS-CFT-09-007}. {To be submitted to JINST}.

\bibitem{Renker:2002st}
D.~Renker, ``{Properties of avalanche photodiodes for applications in high
  energy physics, astrophysics and medical imaging}'', {\em Nucl. Instrum.
  Meth.} {\bf A486} (2002)
164--169.
%%CITATION = NUIMA,A486,164;%%.
  \href{http://dx.doi.org/10.1016/S0168-9002(02)00696-4}{{\tt
  doi:10.1016/S0168-9002(02)00696-4}}.

\bibitem{et1}
See, for example, the technical reprint from ET Enterprises Ltd, ``The
  determination of photomultiplier temperature coefficients for gain and
  spectral sensitivity using the photon counting technique'',
  \url{http://www.electrontubes.com/pdf/rp081colour.pdf}.

\bibitem{et2}
See, for example, Fig. 29 of the brochure from ET Enterprises Ltd,
  ``Understanding Photomultipliers'',
  \url{http://www.electrontubes.com/Photomultipliers/Understanding.pdf}.

\bibitem{CMS_NOTE_2007-028}
M.~Anfreville {et~al.}, ``Laser monitoring system for the CMS lead tungstate
  crystal calorimeter'', {\em CMS Note} {\bf
  \href{http://cms.cern.ch/iCMS/jsp/openfile.jsp?type=NOTE&year=2007&files=NOT%
E2007_028.pdf}{2007/028}} (2007).

\bibitem{Adzic:2006za}
P.~Adzic {et~al.}, ``Results of the first performance tests of the CMS
  electromagnetic calorimeter'', {\em Eur. Phys. J.} {\bf C44S2} (2006)
1--10.
%%CITATION = EPHJA,C44S1,1;%%.
  \href{http://dx.doi.org/10.1140/epjcd/s2005-02-011-3}{{\tt
  doi:10.1140/epjcd/s2005-02-011-3}}.

\bibitem{1748-0221-3-10-P10007}
P.~Adzic {et~al.}, ``Intercalibration of the barrel electromagnetic calorimeter
  of the CMS experiment at start-up'', {\em JINST} {\bf 3} (2008) P10007.
  \href{http://dx.doi.org/10.1088/1748-0221/3/10/P10007}{{\tt
  doi:10.1088/1748-0221/3/10/P10007}}.

\bibitem{Daskalakis:2006dx}
G.~Daskalakis, ``{CMS ECAL calibration strategy}'', {\em AIP Conf. Proc.} {\bf
  867} (2006) 400--407. \href{http://dx.doi.org/10.1063/1.2396978}{{\tt
  doi:10.1063/1.2396978}}.

\bibitem{CMS_NOTE_2009-014}
R.~M. Brown, ``The variation in response of the CMS ECAL vacuum phototriodes as
  a function of orientation in a strong magnetic field'', {\em CMS Note} {\bf
  \href{http://cms.cern.ch/iCMS/jsp/openfile.jsp?type=NOTE&year=2009&files=NOT%
E2009_014.pdf}{2009/014}} (2009).

\end{thebibliography}\endgroup
\cleardoublepage\appendix\section{The CMS Collaboration \label{app:collab}}\begin{sloppypar}\hyphenpenalty=500\textbf{Yerevan Physics Institute,  Yerevan,  Armenia}\\*[0pt]
S.~Chatrchyan, V.~Khachatryan, A.M.~Sirunyan
\vskip\cmsinstskip
\textbf{Institut f\"{u}r Hochenergiephysik der OeAW,  Wien,  Austria}\\*[0pt]
W.~Adam, B.~Arnold, H.~Bergauer, T.~Bergauer, M.~Dragicevic, M.~Eichberger, J.~Er\"{o}
% ...
\vskip\cmsinstskip
% ...

\end{sloppypar}
\end{document}